\documentclass[longbibliography, aps, prx, amsmath, amssymb, amsfonts, twocolumn,
superscriptaddress, footinbib]{revtex4-2}

\usepackage{graphicx}
\usepackage{subfigure}
\usepackage{dcolumn}
\usepackage{bm}
\usepackage{bbm}
\usepackage{amsthm}
\usepackage{mathtools}
\usepackage{physics}
\usepackage{binarytree}
\usepackage{tikz}
\usepackage{verbatim}
\usepackage{pgfplots}
\usepackage{placeins}
\usetikzlibrary{decorations.pathreplacing}
\usetikzlibrary{arrows.meta}
\usetikzlibrary{graphs}

\newcommand{\arrowIn}{
\tikz \draw[-{Stealth[length=2mm, width=1.5mm]}] (-1pt,0) -- (1pt,0);
}
\newcommand{\arrowOut}{
\tikz \draw[-{Stealth[length=2mm, width=1.5mm]}] (1pt,0) -- (-1pt,0);
}

\newcommand{\qd}{\text{d}}
\newcommand{\tTr}{\widetilde{\operatorname{Tr}}}

\usepackage[colorlinks, linkcolor=red, anchorcolor=blue, citecolor=green]{hyperref}

\pgfplotsset{compat=1.16}
\begin{document}

\title[Superactivation of Bell nonlocality in pure anyonic states]
{Superactivation of Bell nonlocality in pure anyonic states}

\author{Cheng-Qian Xu} \email[]{chengqianxu@mail.tsinghua.edu.cn}

\affiliation{State Key Laboratory of Low Dimensional Quantum Physics,
Department of Physics, Tsinghua University, Beijing 100084, China}

\author{Wenhao Ye}

\affiliation{State Key Laboratory of Low Dimensional Quantum Physics,
Department of Physics, Tsinghua University, Beijing 100084, China}

\author{Li You}

\affiliation{State Key Laboratory of Low Dimensional Quantum Physics,
Department of Physics, Tsinghua University, Beijing 100084, China}

\affiliation{Frontier Science Center for Quantum Information, Beijing, China}

\affiliation{Beijing Academy of Quantum Information Sciences, Beijing 100193, China}

\affiliation{Hefei National Laboratory, Hefei, Anhui 230088, China}

\date{\today}

\begin{abstract}
    Standard quantum information theory is founded on the assumption that multi-party state space
    possesses a tensor product structure.
    Anyons, as quasiparticles in two-dimensional systems, exhibit unique entanglement properties
    that differ from the conventional quantum systems, resulting from the absence of
    a tensor product structure in their state spaces.
    This motivates us to investigate the relationship between Bell nonlocality and entanglement
    in anyonic states.
    Specifically, we find that certain pure anyonic states with non-zero anyonic entanglement
    entropy (AEE) are local, yet exhibit nonlocality when subjected to collective
    measurements on multiple copies—a phenomenon known as superactivation of nonlocality,
    which is typically observed in conventional mixed states.
    To analyze this, we decompose the total entanglement of anyonic states
    into two components: one from the tensor product structure and the other
    representing residual contributions.
    By studying their asymptotic behavior, we find that the former gradually increases
    and approaches the AEE while the latter diminishes with the number of copies.
    Crucially, the entanglement component associated with the tensor product structure
    demonstrates a significant correlation with nonlocality,
    which explains the observed superactivation of nonlocality.
    Our findings provide new insights into the connection between entanglement and nonlocality
    in anyonic systems.
\end{abstract}

                              
\maketitle

\section{Introduction}

Anyons~\cite{PhysRevLett.49.957, PhysRevLett.48.1144} or particles whose statistical behavior
generalizes the concept of bosons and fermions, have attracted significant attention in the fields
of quantum computing and condensed matter physics~\cite{kitaev2006anyons, RevModPhys.80.1083, RN8,
doi:10.1126/science.aaz5601, Aghaee2025-el}.
Unlike conventional quantum systems, anyonic systems lack a tensor product structure in their state
spaces, resulting in a fundamentally different entanglement phenomenon known as topological entanglement
entropy, which arises specifically in many-body systems~\cite{kitaev2006topological, levin2006detecting,
PhysRevLett.131.166601, PhysRevB.110.165154}.
Furthermore, within quantum information theory, significant research efforts are dedicated
to characterizing this unique form of entanglement called
anyonic charge entanglement~\cite{bonderson2008, bonderson2017anyonic, PhysRevA.108.052221, YXY},
whose unique property raises intriguing questions about the relation between Bell nonlocality
and entanglement in such systems.

Bell nonlocality is a core concept in quantum mechanics, describing correlations between parties or
quantum systems that cannot be explained by locality~\cite{PhysicsPhysiqueFizika.1.195, RevModPhys.86.419}.
The Bell nonlocality of quantum states has been experimentally demonstrated through violations of
Bell inequalities~\cite{PhysRevLett.23.880, PhysRevLett.28.938, PhysRevLett.49.1804, PhysRevLett.49.91,
PhysRevLett.81.5039, Hensen2015-cp, PhysRevLett.121.080404}.
In conventional quantum information theory,  it is known that a pure state is local
iff it has zero entanglement entropy~\cite{GISIN199215}.
However, in anyonic systems, the situation is more complex due to the non-tensor product nature
of their state spaces.
This complexity motivates us to investigate whether pure anyonic states may exhibit phenomena
that are not observed in conventional pure states.

In this paper, we study the relation between Bell nonlocality and entanglement in pure
anyonic states.
Within the framework of our proposed Bell nonlocality detection scheme for multi-copy anyonic states,
we find that the anyonic entanglement entropy (AEE)~\cite{hikami2008skein, PhysRevB.89.035105,
PhysRevA.90.062325, bonderson2017anyonic}
is not a sufficient condition for Bell nonlocality of pure anyonic states,
instead, we find that there exists pure anyonic state with non-zero AEE but is nevertheless local.
Using collective measurements on multiple copies of anyonic states for Bell inequality test,
we discover that these local states miraculously can exhibit Bell nonlocality.
This phenomenon, known before as superactivation of nonlocality~\cite{PhysRevLett.109.190401},
is a unique feature of conventional mixed states.

Our analysis is based on the decomposition formula [see Eq. (\ref{eq:Pytha})]
for anyonic entanglement established in Ref.~\cite{YXY}.
The total entanglement of anyonic states is decomposed into two distinct components:
one arising from the tensor product structure in the anyonic states spaces and the other from
the non-tensor product structure.
By analyzing the asymptotic behavior of these two entanglement components, we discover how their
sizes change as the number of copies increases and where they ultimately converge to,
providing an intuitive understanding of their interplay.
Because Bell experiments involve the measurement of the tensor product of local observables,
the entanglement components (rather than the AEE) originating from the anyonic state spaces satisfying
the tensor product structure serve to characterize Bell nonlocality, which we believe
is directly responsible for the observed superactivation of nonlocality in pure anyonic states.

It is noteworthy to point out that similar phenomena have also been investigated through the lens of
superselection rules in optics or other systems governed by conservation
laws~\cite{PhysRevA.73.022311, PhysRevLett.92.180401, PhysRevLett.92.193601, RevModPhys.79.555}.
However, anyonic system we study here is relatively more complex.
First, the methods involved in previous studies~\cite{PhysRevLett.91.097902, PhysRevLett.91.097903}
cannot be directly applied to anyonic systems because anyonic superselection rules cannot be represented
by a compact symmetry group, which is different from other superselection rules~\cite{PhysRevA.69.052326}.
Second, anyonic systems are further complicated by the presence of non-Abelian fusion rules,
which enrich the structure of their state spaces with added complexity.

This paper is organized as follows.
In Sec.~\ref{sec:2}, we briefly review the Fibonacci anyon model and provide a definition of pure
anyonic states from an information-theoretic perspective.
Our conclusions can be readily generalized to other anyon models.
In Sec.~\ref{sec:3}, we present an operational scheme for testing Bell inequalities of multi-copy
anyonic states.
We demonstrate that the pure anyonic state $\ket{\tau, \tau; 1}$ exhibits superactivation
of nonlocality.
Then, in Sec.~\ref{sec:4}, we decompose the total entanglement of anyonic states into two components, and
investigate the asymptotic behavior of anyonic entanglement to elucidate the superactivation of nonlocality.
In Sec.~\ref{sec:5}, we demonstrate the relationship between entanglement and Bell nonlocality
in pure anyonic states.
Finally we end with a summary.

\section{\label{sec:2} Fibonacci Anyon Model and Pure Anyonic States}

In this section, we will review the Fibonacci anyon
model~\cite{kitaev2006anyons, trebst2008short, pachos_2012, 2001Lectures} and establish
the definition of pure anyonic states from an information-theoretic perspective.

\subsection{Fibonacci Anyon Model}

The Fibonacci anyon model contains two distinct topological charges:
 $\left\{ 1, \tau \right\}$,
where $1$ represents the vacuum and $\tau$ denote the Fibonacci anyon.
These anyons obey the fusion rules:
\begin{align}
    & 1 \times \tau = \tau, \nonumber \\
    & \tau \times 1 = \tau, \nonumber \\
    & \tau \times \tau = 1 + \tau. \nonumber
\end{align}
The vacuum $1$ is trivial since the result of any other anyon fusing with the vacuum
gives itself, and there are two possible fusion results when two $\tau$s are fused.
For a given anyon $a$, there exists an unique anyon $\overline{a}$ such that the result of their
fusion contains a vacuum. Here, $\overline{\tau} = \tau$.

The Hilbert space construction follows from these fusion rules. For two $\tau$ anyons,
the resulting two-dimensional Hilbert space is spanned by the orthonormal basis:
\begin{align}
    \label{eq:1}
    & \ket{\tau, \tau; 1} = \left(\frac{1}{\qd_\tau}\right)^{\frac{1}{2}}
    \begin{tikzpicture}
        \draw (-0.5,0.25) -- (0,-0.25) -- (0.5,0.25);
    \end{tikzpicture}, \nonumber \\
    & \ket{\tau, \tau; \tau} = \left(\frac{1}{\qd_\tau}\right)^{\frac{1}{4}}
    \begin{tikzpicture}[baseline]
        \draw (-0.5,0.5) -- (0,0) -- (0.5,0.5);
        \draw (0,0) -- (0,-0.5);
    \end{tikzpicture},
\end{align}
where $\qd_\tau = \frac{\sqrt{5}+1}{2}$ in normalization coefficients is the quantum
dimension of $\tau$ anyon.
We have used the diagrammatic representation of anyon model.
Each anyon is associated with an oriented (the arrow is omitted here) line going up
from the bottom.
Here, $\tau$ is represented by the solid line and the vacuum is omitted.
For cases with more than two $\tau$s, we need to specify the order of fusion
since different orders will give different bases of the same Hilbert space.
These different bases can be transformed into each other via natural isomorphic
transformations named $F$-matrices.

In addition to the fusion rules above, Fibonacci anyons are also restricted by
the braiding rules.
Specifically, exchanging two anyons will give a unitary transformation,
called the $R$-matrices, acting on the Hilbert space where these two anyons live.
Through braiding operators, Fibonacci anyon model was shown to be able to realize
any unitary gate with any precision.
This anyon model is therefore expected to enable universal quantum
computing~\cite{CMP2002Freedman, preskill1999lecture}.

In the operator space based on anyonic Hilbert space, Quantum trace $\tTr$,
which generalizes the concept of trace $\operatorname{Tr}$, is defined by connecting
the outgoing lines and the corresponding incoming lines of the operator, for example,
\begin{equation}
    \tTr \left( \ket{\tau, \tau, 1} \bra{\tau, \tau; 1} \right)
        = \frac{1}{\qd_\tau}
        \begin{tikzpicture}[baseline]
            \draw (-0.5,0.75) -- (0,0.25) -- (0.5,0.75) -- (1,0.25) -- (1,-0.25);
            \draw (-0.5,-0.75) -- (0,-0.25) -- (0.5,-0.75) -- (1,-0.25);
            \draw (-0.5,0.75) -- (-1,0.25) -- (-1,-0.25) -- (-0.5,-0.75);
        \end{tikzpicture} = 1, \nonumber
\end{equation}
where a loop of $\tau$ is defined as the quantum dimension $\qd_\tau$, which is also
the orign of the normalization coefficients in Eq.~(\ref{eq:1}).
The anyonic state space, a subspace of the operator space, encompasses all anyonic density
matrices $\tilde{\rho}$, which is positive semi-definite $\tilde{\rho} \ge 0$ and
normalized $\tTr \left( \tilde{\rho} \right) = 1$.
The anyonic state space lacks a tensor product structure, meaning there are bipartite anyonic
states $\tilde{\rho}_{AB}$ that cannot be expressed in the form of
\begin{equation}
    \tilde{\rho}_{AB} \neq \sum_{i,j} \alpha_{ij} \sigma_A^i \otimes \sigma_B^j, \nonumber
\end{equation}
where $\left\{\alpha_{ij}\right\}_{i,j}$ represent the coefficients,
meanwhile, $\left\{ \sigma_A^i \right\}_i$ and $\left\{ \sigma_B^j \right\}_j$ denote
complete sets of observables for subsystem $A$ and $B$, respectively~\cite{PhysRevA.108.052221}.

In addition, we will use the so-called $\Omega$-loop,
\begin{equation}
    \begin{tikzpicture}[baseline]
        \draw[line width=2pt] (0.4,0) arc(0:360:0.4 and 0.2); \draw (-0.4,0) node[left]{$\Omega$};
    \end{tikzpicture} = \sum_{a = \left\{1,\tau\right\}} \frac{\qd_a}{\mathcal{D}^2}
    \begin{tikzpicture}[baseline]
        \draw[] (0.4,0) node[right]{$a$} arc(0:360:0.4 and 0.2);
    \end{tikzpicture}, \nonumber
\end{equation}
where $\mathcal{D} = \sqrt{\qd_1^2 + \qd_\tau^2}$. $\Omega$-loop allows only
the vacuum to pass through it:
\begin{equation}
    \begin{tikzpicture}[baseline]
        \draw[line width=2pt] (-0.07,0.2) arc(100:440:0.4 and 0.2);
        \draw (-0.4,0) node[left]{$\Omega$};
        \draw (0,-0.07) -- (0,0.7) node[right]{$a$}; \draw (0,-0.3) -- (0,-0.5);
    \end{tikzpicture} = \delta_{1a}, \nonumber
\end{equation}
for $a \in \left\{1, \tau\right\}$.

\subsection{Pure Anyonic States}

Current debates regarding purity of anyonic state are found in Ref.~\cite{Vidal2024}.
The controversy arises when states like $\ket{\tau,\tau; \tau}$ with non-Abelian
total charge are considered.
While computational frameworks treat $\ket{(\tau,\tau;1),\tau;\tau}$
and $\ket{(\tau,\tau;\tau),\tau;\tau}$ as pure states,
the quantum trace condition $\widetilde{\operatorname{Tr}}[\tilde{\rho}^2] = 1$ reveals their
mixed nature under generalized purity criteria,
the latter of which is supported from an information-theoretic perspective.
More specifically, our definition of pure anyonic states is given in the following.

\textbf{Definition~1 (Pure Anyonic State)} --
An anyonic state $\tilde{\rho}$ is pure iff its anyonic entropy is zero
\begin{equation}
    \widetilde{S}(\tilde{\rho}) = - \tTr \left( \tilde{\rho} \log_2 \tilde{\rho} \right) = 0,
    \nonumber
\end{equation}
otherwise, it is a mixed anyonic state.

This entropy-based definition preserves the operational interpretation of
von Neumann entropy.
It is known that the von Neumann entropy of a quantum state is the minimal and reliable
quantity of information needed to store the exact state of the system asymptotically
in the most compressed form~\cite{nielsen2002quantum, PhysRevA.51.2738}.
Pure state is deterministic, therefore the amount of information to store a pure state
is zero.
It is expected that this property to be maintained in anyonic systems as well.

Now we consider anyonic state $\ket{\tau,\tau; \tau}$.
If it is pure, then the minimal and reliable amount of information to store it
in the asymptotic limit should be zero.
However, due to the non-Abelian fusion rule of Fibonacci anyons, two copies
of $\tilde{\rho} = \frac{1}{\qd_\tau} \ket{\tau,\tau; \tau} \bra{\tau,\tau;\tau}$
span the two-dimensional Hilbert space,
\begin{align}
     \frac{1}{\qd_\tau^3} 
    \begin{tikzpicture}[baseline]
        \draw (-0.5,0.75) -- (0,0.25) -- (0.5,0.75);
        \draw (0,0.25) -- (0,-0.25);
        \draw (-0.5,-0.75) -- (0,-0.25) -- (0.5,-0.75);
    \end{tikzpicture} \otimes
    \begin{tikzpicture}[baseline]
        \draw (-0.5,0.75) -- (0,0.25) -- (0.5,0.75);
        \draw (0,0.25) -- (0,-0.25);
        \draw (-0.5,-0.75) -- (0,-0.25) -- (0.5,-0.75);
    \end{tikzpicture} 
    =   \frac{1}{\qd_\tau^4}
    \begin{tikzpicture}[baseline]
        \draw (-0.5,0.75) -- (0,0.25) -- (0.5,0.75);
        \draw (-0.5,-0.75) -- (0,-0.25) -- (0.5,-0.75);
        \draw (-0.1,0.75) -- (-0.3,0.55);
        \draw (0.1,0.75) -- (0.3,0.55);
        \draw (-0.1,-0.75) -- (-0.3,-0.55);
        \draw (0.1,-0.75) -- (0.3,-0.55);
    \end{tikzpicture} + \left( \frac{1}{\qd_\tau} \right)^{\frac{7}{2}}
    \begin{tikzpicture}[baseline]
        \draw (-0.5,0.75) -- (0,0.25) -- (0.5,0.75);
        \draw (-0.5,-0.75) -- (0,-0.25) -- (0.5,-0.75);
        \draw (-0.1,0.75) -- (-0.3,0.55);
        \draw (0.1,0.75) -- (0.3,0.55);
        \draw (-0.1,-0.75) -- (-0.3,-0.55);
        \draw (0.1,-0.75) -- (0.3,-0.55);
        \draw (0,0.25) -- (0,-0.25);
    \end{tikzpicture}. \nonumber
\end{align}
In the asymptotic limit, the average quantity of information needed
to store $\ket{\tau,\tau;\tau}$ is equal to
anyonic entropy $\widetilde{S}(\ket{\tau,\tau;\tau}) = \log_2 \qd_\tau$.
This makes sense because the non-Abelian total charge carries information that is not
locally accessible, while the state seems pure, this part of the information
becomes accessible only through fusion rules in the multi-copy setting.
Thus, anyonic state $\ket{\tau,\tau;\tau}$ is not pure in this sense.

Pure anyonic states given by Definition~1 inherit the entanglement symmetry property
possessed by conventional pure states.
The Schmidt decomposition for a bipartite pure anyonic state takes the form:
\begin{equation}\label{eq:SchPS}
    \ket{\tilde{\psi}}_{AB}
    = \sum_{c \in \left\{ 1, \tau \right\}} \sum_{i=1}^{K_c}
        \sqrt{\lambda_{c,i}} \ket{i; c}_A \otimes \ket{i; \overline{c}}_B \otimes
        \ket{c, \overline{c}; 1}_F,
\end{equation}
where $\left\{ \ket{i; c}_A \right\}$ and $\left\{ \ket{i; \overline{c}}_B \right\}$
are the canonical bases of the Hilbert spaces of subsystems $A$ and $B$, respectively,
the subscript $F$ indicates the fusion space
beyond the subsystems $A$ and $B$, $K_c$ denotes the Schmidt rank of sector-$c$,
and $\lambda_{c,i}$ are non-negative real numbers called Schmidt coefficients
that satisfy $\sum_{c,i} \lambda_{c,i} = 1$.
Its entanglement can be described by the AEE,
\begin{equation}
    E_{\rm AEE} = \widetilde{S}(\tilde{\rho}_A) = \widetilde{S}(\tilde{\rho}_B)
        = H(\left\{ \lambda_{c,i} \right\}) + p_\tau \log_2 \qd_\tau, \nonumber
\end{equation}
where $H(\left\{ \lambda_{c,i} \right\}) = \sum_{c,i} - \lambda_{c,i} \log_2 \lambda_{c,i}$
is the Shannon entropy, $p_\tau = \sum_i \lambda_{\tau, i}$, $\tilde{\rho}_A$
and $\tilde{\rho}_B$ are anyonic reduced density matrices of $\ket{\tilde{\psi}}_{AB}$.
In the above discussion, we only consider the Fibonacci anyon model, in which vacuum is
the only Abelian charge.
The case where the total charge of the anyonic state is any other Abelian charge
can be found in Appendix~\ref{app:Sym}.

\section{\label{sec:3} Bell Inequality Violation with Pure Anyonic States}

In the following, we investigate the violation of Bell inequality for pure anyonic
states.
Specifically, some of the anyons that comprise of the pure anyonic state $\tilde{\rho}_{AB}$
are distributed to Alice and the rest to Bob.
Each of them performs local measurements to obtain possible outcomes,
labeled by $a \in \left\{ 1,\cdots, K \right\}$ for Alice
and $b \in \left\{ 1, \cdots, K \right\}$ for Bob, of some observables,
labeled by $x \in \left\{ 1, \cdots, J \right\}$ for Alice
and $y \in \left\{ 1, \cdots, J \right\}$ for Bob.
After repeating the above test many times, they calculate the probability
distribution $P_{AB}(a,b|x,y)$ for observing $x$ and $y$ with results $a$
and $b$, respectively.
Formally, the probability distribution $P_{AB}(a,b|x,y)$ can be written as
\begin{equation}
    P(a,b|x,y) = \tTr \left( \tilde{\rho}_{AB} M_x^a \otimes M_y^b \right), \nonumber
\end{equation}
where $M_x^a$ ($M_y^b$) are positive operator valued measurement (POVM) elements
characterizing Alice's (Bob's) measurements.
We say that collection $\left\{ P_{AB}(a,b|x,y) \right\}$ or the anyonic
state $\tilde{\rho}_{AB}$ is nonlocal if some probability distribution $P_{AB}(a,b|x,y)$
cannot be described by a local hidden variable model (LHVM)
\begin{equation}
    P_{AB}(a,b|x,y) = \int_{\Omega} d\mathbb{P}(\lambda) q(\lambda) P_A(a|x,\lambda)
        P_B(b|y,\lambda), \nonumber
\end{equation}
where $\left( \Omega, \mathcal{F}, \mathbb{P} \right)$ defines a probability space, $\lambda$
is hidden variable, and $q(\lambda)$ is its probability density.
Alternatively, the anyonic state $\tilde{\rho}_{AB}$ is nonlocal if it violates certain
Bell inequalities.
Here, we consider Clauser-Horne-Shimony-Holt (CHSH) inequality
\begin{equation}
    \label{eq:CHSH}
    |\langle A_1 B_1 \rangle + \langle A_1 B_2 \rangle + \langle A_2 B_1 \rangle
        - \langle A_2 B_2 \rangle| \le 2, \nonumber
\end{equation}
where $\langle A_x B_y \rangle = \tTr \left( \tilde{\rho} A_x B_y\right)$
for $x, y \in \left\{ 1,2 \right\}$, and $A_x$ ($B_y$) are Hermitian operators
with eigenvalues $+1$ and $-1$ characterizing Alice's (Bob's) observables.

It is known that a conventional pure
state $\ket{\psi}_{AB} = \sum_{i=1}^K \sqrt{\lambda_i} \ket{i}_A \ket{i}_B$ is nonlocal
if and only if it is entangled, or, equivalently, with Schmidt rank $K > 1$.
For pure anyonic states, however, this conclusion no longer holds.
Remarkably, the entangled anyonic state
\begin{equation}
    \ket{\tau, \tau; 1} = \frac{1}{\sqrt{\qd_\tau}}
    \begin{tikzpicture}[baseline]
        \draw (-0.5,0.5) -- (0,0) -- (0.5,0.5);
    \end{tikzpicture}, \nonumber
\end{equation}
with nontrivial AEE
\begin{equation}
    E_{\rm AEE}(\ket{\tau, \tau; 1}) = \log_2 \qd_\tau, \nonumber
\end{equation}
exhibits trivial local hidden variable behavior for the charge $\tau$ as the hidden
variable
\begin{equation}
    P_{AB}(\tau, \tau|x, y) = P_A(\tau|x, \tau) P_B(\tau|y, \tau), \nonumber
\end{equation}
where $x$ and $y$ are local observables that depend on the charge of anyons.
This contrasts with the conventional pure quantum systems where entanglement
implies nonlocality.

It is permissible for some conventional quantum mixed states to be entangled without
violating Bell inequality in quantum information theory~\cite{PhysRevA.40.4277,
PhysRevA.65.042302}, e.g., isotropic states
in a certain parameter range~\cite{PhysRevLett.99.040403, PhysRevLett.98.140402}.
These quantum mixed states may lead to nonlocality for some more complex
scenarios in a subtle way.
For instance, joint measurements on several copies of a local state have been
considered for revealing nonlocality, which is known as superactivation.
We are thus inspired to carry out this study to answer whether superactivation of
nonlocality exists in pure anyonic states.

\subsection{Multiple Copies of an Anyonic State}

Before exploring superactivation of nonlocality in anyonic systems,
it is essential to establish a rigorous definition for multiple copies of a bipartite
anyonic state within quantum information protocols.
In conventional quantum systems, two-copy of a bipartite state $\rho_{AB}$ naturally
follows the tensor product structure:
\begin{equation}
    \label{eq:Multiple}
    \rho_{AB}^{\otimes 2} = \rho_{AB} \otimes \rho_{AB},
\end{equation}
where Alice controls all subsystems $A$ while Bob governs subsystems $B$, with their
respective local operations maintaining strict causal separation.
For anyonic systems, it is natural to extend the form in Eq.~(\ref{eq:Multiple}),
for example, two-copy of the state $\ket{\tau, \tau; 1}$ expressed as below
\begin{equation}
    \label{eq:Multiple1}
    \ket{\tau, \tau; 1}^{\otimes 2}
        = \frac{1}{\qd_\tau}
        \begin{tikzpicture}[baseline]
            \draw (0.25,0.5) node[above]{$A_2$} -- (0.75,0) -- (1.25,0.5) node[above]{$B_2$};
            \draw (-0.25,0.5) node[above]{$B_1$} -- (-0.75,0) -- (-1.25,0.5) node[above]{$A_1$};
        \end{tikzpicture}.
\end{equation}
Subsystem ownership is explicitly labeled in this diagrammatic representation.
When Alice attempts to perform local operations via braiding Fibonacci anyons $A_1$ and $A_2$,
a problem arises as she has to consider the interaction of her operations with $B_1$ and $B_2$.
Similarly, Bob faces nontrivial challenges when he attempts to discard subsystem $B_1$.
These problems can be solved by redefining the concept of multiple copies of
an anyonic state. Consider the following revised two copies of $\ket{\tau, \tau; 1}$:
\begin{equation}
    \label{eq:Multiple2}
    \ket{\tau, \tau; 1}^2 = \frac{1}{\qd_\tau}
        \begin{tikzpicture}[baseline]
            \draw (-0.5,0.5) node[above]{$A_2$} -- (0,0) -- (0.5,0.5) node[above]{$B_2$};
            \draw (-1,0.5) node[above]{$A_1$} -- (0,-0.5) -- (1,0.5) node[above]{$B_1$};
        \end{tikzpicture}.
\end{equation}
Crucially, states (\ref{eq:Multiple1}) and (\ref{eq:Multiple2}) are physically equivalent
in describing two pairs of $\ket{\tau, \tau; 1}$ differing only in the numbering of orders.
As demonstrated in Ref.~\cite{PhysRevA.90.062325}, this relation permits embedding the anyonic
state space of Eq.~(\ref{eq:Multiple1}) into the anyonic state space of Eq.~(\ref{eq:Multiple2})
by introducing the larger (non-physical) Hilbert space -- a mathematical framework
enabling investigation of anyonic entanglement under asymptotic limit.
This approach requires the total topological charge of a single anyonic state is Abelian.

Here, we give an explicit operational transformation between states Eqs.~(\ref{eq:Multiple1})
and (\ref{eq:Multiple2}).
Let's define the tangled braiding $\mathcal{T}$~\cite{PhysRevA.106.012413}:
\begin{equation}
    \mathcal{T} = \begin{tikzpicture}[baseline]
        \draw (0,1) -- (0,-0.5);
        \draw (0.5,-0.5) -- (1.5,0.5) -- (1.5,1);
        \draw (0.8,-0.3) -- (1,-0.5);
        \draw (1.5,-0.5) -- (1.5,0) -- (1.3,0.2);
        \draw (0.7,-0.2) -- (0.5,0) -- (0.5,0.5) -- (0.5,1);
        \draw (1.2,0.3) -- (1,0.5) -- (1,1);
    \end{tikzpicture}, \nonumber
\end{equation}
which can transform the state $\ket{\tau, \tau; 1}^{\otimes 2}$ into the
state $\ket{\tau, \tau; 1}^2$.
It can be easily generalized to general anyonic
state $\tilde{\rho}_{AB}$ such that $\tilde{\rho}_{AB}^n = \mathcal{T}
 \tilde{\rho}_{AB}^{\otimes n} \mathcal{T}^\dagger$ (see Appendix~\ref{app:TB}).

The advantage of employing the tangled braiding $\mathcal{T}$ lies in its ability to
arrange multiple copies of $\tilde{\rho}_{AB}^{\otimes n}$ such that Alice's local
operations will not influence Bob's subsystems. Specifically, we have
\begin{equation}
    \label{eq:LocalEq}
    \tTr_A \left[ \mathcal{E}_A( \mathcal{T} \tilde{\rho}_{AB}^{\otimes n} \mathcal{T}^\dagger) \right]
        = \tilde{\rho}_B^{\otimes n},
\end{equation}
where $\mathcal{E}(\tilde{\sigma})$ denotes an arbitrary trace-preserving quantum operation
acting on anyonic state $\tilde{\sigma}$,
and $\tilde{\rho}_B = \tTr_A \left( \tilde{\rho}_{AB} \right)$.
Using Eq.~(\ref{eq:LocalEq}), we can obtain an equality:
\begin{equation}
    \label{eq:TensorAEE}
    E_{\rm AEE} (\mathcal{T}\tilde{\rho}_{AB}^{\otimes n} \mathcal{T}^\dagger)
        = n E_{\rm AEE}(\tilde{\rho}_{AB}),
\end{equation}
indicating that after applying $\mathcal{T}$, the AEE of an $n$-copied system
becomes exactly $n$ times the AEE of a single copy.
Notably, The operator $\mathcal{T}$ gives the definition of AEE for anyonic
state $\tilde{\rho}^{\otimes n}$.
The proofs of Eqs. (\ref{eq:LocalEq}) and (\ref{eq:TensorAEE})
are provided in Appendix~\ref{app:TB}.

This framework justifies defining the $n$-copy of an anyonic
state $\tilde{\rho}_{AB}$ as $\tilde{\rho}_{AB}^n$ which can be operationally
realized by applying $\mathcal{T}$ to $\tilde{\rho}_{AB}^{\otimes n}$.
In essence, the tangled braiding $\mathcal{T}$ is merely a protocol convention designed to
ensure that, under such an arrangement, local operations within subsystems do not
interfere with each another.

\subsection{Superactivation of Nonlocality}

In the following we demonstrate the superactivation of quantum nonlocality in anyonic systems
by investigating the Bell inequality test of the three-copy of $\ket{\tau, \tau; 1}$.
For two-copy,
\begin{equation}
    \ket{\tau, \tau; 1}^2 = \frac{1}{\qd_\tau^2}
    \begin{tikzpicture}[baseline]
        \draw (0.25,0.5) node[above]{$B_2$} -- (0.5,0.25) -- (0.75,0.5) node[above]{$B_1$};
        \draw (-0.25,0.5) node[above]{$A_2$} -- (-0.5,0.25) -- (-0.75,0.5) node[above]{$A_1$};
    \end{tikzpicture}
    + \left( \frac{1}{\qd_\tau} \right)^{\frac{3}{2}}
    \begin{tikzpicture}[baseline]
        \draw (0.25,0.5) node[above]{$B_2$} -- (0.5,0.25) -- (0.75,0.5) node[above]{$B_1$};
        \draw (-0.25,0.5) node[above]{$A_2$} -- (-0.5,0.25) -- (-0.75,0.5) node[above]{$A_1$};
        \draw (-0.5,0.25) -- (0,-0.25) -- (0.5,0.25);
    \end{tikzpicture}, \nonumber
\end{equation}
Alice and Bob cannot transform the state with charge-$1$, $\ket{\tau, \tau; 1}$, to
the state with charge-$\tau$, $\ket{\tau, \tau; \tau}$, due to the superselection rules
for topological charges~\cite{PhysRevA.69.052326}.
Therefore, anyonic state $\ket{\tau, \tau; 1}^2$ is local and can be described by
LHVM with the total charge of the subsystem as the hidden variable,
even though it is entangled with non-zero AEE.

The Schmidt decomposition for three-copy of $\ket{\tau, \tau; 1}$ takes the form:
\begin{align}
    & \ket{\tau, \tau; 1}^3  \nonumber \\
    = & \left( \frac{1}{\qd_\tau} \right)^{\frac{3}{2}}
        \ket{(\tau,\tau)_\tau, \tau; 1}_A \otimes \ket{(\tau, \tau)_{\tau}, \tau; 1}_B \nonumber \\
    & + \frac{1}{\qd_\tau} \sum_{i = 1, \tau} \ket{(\tau,\tau)_i, \tau; \tau}_A \otimes
        \ket{(\tau, \tau)_i, \tau; \tau}_B \otimes \ket{\tau, \tau; 1}_F, \nonumber 
\end{align}
where Alice and Bob each have the ($1+2$)-dimensional graded Hilbert
space $\mathcal{H}_1 \oplus \mathcal{H}_\tau$ due to the superselection rules,
spanned by the canonical base
\begin{equation}
    \label{eq:base}
    \left\{ \ket{(\tau, \tau)_\tau, \tau; 1} \right\} \oplus \left\{ \ket{(\tau, \tau)_1, \tau; \tau},
        \ket{(\tau, \tau)_\tau, \tau; \tau} \right\}.
\end{equation}
The Bell nonlocality is expected to be activated in the Hilbert subspace $\mathcal{H}_\tau$.

In this Hilbert space $\mathcal{H}_1 \oplus \mathcal{H}_{\tau}$,
the Hermitian operator (with eigenvalues $\pm 1$) employed by Alice can be parameterized
in the canonical basis (\ref{eq:base})
\begin{equation}
    A_\alpha = \begin{pmatrix}
        1 & 0 & 0 \\
        0 & \cos \alpha & \sin \alpha \\
        0 & \sin \alpha & -\cos \alpha \\ 
    \end{pmatrix}.
\end{equation}
where $\alpha$ is a parameter that takes values from $0$ to $2\pi$.
Analogously, Bob's measurement operator $B_\beta$ adopts an identical parametrization
with $\beta$ replacing $\alpha$.
By substituting the aforementioned $A_\alpha$ and $B_\beta$ into the CHSH
quantity (\ref{eq:CHSH}) for pure anyonic state $\ket{\tau, \tau; 1}^3$, we can get
\begin{align}
    & \langle A_{\alpha_1} B_{\beta_1} \rangle + \langle A_{\alpha_1} B_{\beta_2} \rangle
        + \langle A_{\alpha_2} B_{\beta_1} \rangle - \langle A_{\alpha_2} B_{\beta_2} \rangle \nonumber \\
    = & \frac{2}{\qd_\tau^2} \big[ \cos(\alpha_1 - \beta_1) + \cos(\alpha_1 - \beta_2)
        + \cos(\alpha_2 - \beta_1) \nonumber \\
    & - \cos(\alpha_2 - \beta_2) \big] + \frac{2}{\qd_\tau^3} \nonumber \\
    \le & \frac{4\sqrt{2} \qd_\tau + 2}{\qd_\tau^3} \approx 2.63286
\end{align}
where the equality is obatined for $\alpha_1 = 0$, $\alpha_2 = \frac{\pi}{2}$,
$\beta_1 = \frac{\pi}{4}$ and $\beta_2 = - \frac{\pi}{4}$, which violates the CHSH inequality.

\section{\label{sec:4} The Asymptotic Behavior of Anyonic Entanglement}

How can we interpret the superactivation of nonlocality in pure anyonic states from
an information-theoretic perspective, a phenomenon once thought exclusive to conventional
mixed systems?
In this section, we systematically examine the asymptotic behavior of anyonic entanglement
to uncover its intrinsic nonlocal properties.

In Ref.~\cite{YXY}, based on the anyonic relative entropy $\widetilde{S}( \tilde{\rho} \| \tilde{\sigma})
 = \tTr \left( \tilde{\rho} \log_2 \tilde{\rho} \right) - \tTr \left( \tilde{\rho} \log_2
 \tilde{\sigma} \right)$,
the authers proposed a formula for entanglement in anyonic state $\tilde{\rho}$
\begin{equation}
    \label{eq:Pytha}
    \min_{\tilde{\sigma} \in {\rm SEP}} \widetilde{S} (\tilde{\rho} \| \tilde{\sigma})
        = \widetilde{S}(\tilde{\rho} \| \Omega(\tilde{\rho})) + \min_{\tilde{\sigma} \in {\rm SEP}}
        \widetilde{S} (\Omega(\tilde{\rho}) \| \tilde{\sigma}),
\end{equation}
where SEP denotes the set of separable anyonic states $\tilde{\rho} = \sum_k p_k \tilde{\rho}_A^k
 \otimes \tilde{\rho}_B^k$, and $\Omega(\tilde{\rho})$ is a mapping that severs all charge lines
between the subsystems of the bipartite anyonic state~\cite{bonderson2017anyonic},
\begin{align}
    & \Omega \left(~ \begin{tikzpicture}[baseline, scale=0.5]
        \draw (-0.5,0.75) node[above]{$a$} -- (0,0.25) -- (0.5,0.75) node[above]{$b$};
        \draw (-0.5,-0.75) node[below]{$a'$} -- (0,-0.25) -- (0.5,-0.75) node[below]{$b'$};
        \draw (0,0.25) -- (0,-0.25) node[pos=0.5, right]{$c$};
    \end{tikzpicture}~ \right)
    = \sum_e \left[ F^{ab}_{a' b'} \right]_{ce} \Omega \left(~ \begin{tikzpicture}[baseline, scale=0.5]
        \draw (-0.75,0.5) node[above]{$a$} -- (-0.25,0) -- (-0.75,-0.5) node[below]{$a'$};
        \draw (0.75,0.5) node[above]{$b$} -- (0.25,0) -- (0.75,-0.5) node[below]{$b'$};
        \draw (-0.25,0) -- (0.25,0) node[pos=0.5, above]{$e$};
    \end{tikzpicture}~ \right) \nonumber \\
    = & \sum_e \left[ F^{ab}_{a' b'} \right]_{ce}
    \begin{tikzpicture}[baseline, scale=0.8]
        \draw (-0.75,0.5) node[above]{$a$} -- (-0.25,0) -- (-0.75,-0.5) node[below]{$a'$};
        \draw (0.75,0.5) node[above]{$b$} -- (0.25,0) -- (0.75,-0.5) node[below]{$b'$};
        \draw (-0.25,0) -- (0.25,0) node[above]{$e$};
        \draw[line width=2pt] (0.1,0.03) arc(10:350:0.1 and 0.2);
        \draw (0,-0.2) node[below]{$\Omega$};
    \end{tikzpicture} = \sqrt{\frac{\qd_c}{\qd_a \qd_b}} \delta_{aa'} \delta_{bb'}
    \begin{tikzpicture}[baseline, scale=0.8]
        \draw (-0.25,-0.5) -- (-0.25,0.5) node[above]{$a$};
        \draw (0.25,-0.5) -- (0.25,0.5) node[above]{$b$};
    \end{tikzpicture}, \nonumber
\end{align}
where latin letters $a,a',b,b',c,e \in \left\{1, \tau\right\}$ for Fibonacci anyon model
and we have suppressed variables within subsystems.
It can be intuitively seen that $\Omega(\tilde{\rho})$ projects the state $\tilde{\rho}$ onto
the anyonic state space that satisfies the tensor product structure.

The left side of Eq.~(\ref{eq:Pytha}) is the anyonic relative entropy of entanglement (AREE),
 $E_{\rm AREE} \equiv \min_{\tilde{\sigma} \in {\rm SEP}}
 \widetilde{S} (\tilde{\rho} \| \tilde{\sigma})$,
which measures the total entanglement of the anyonic state $\tilde{\rho}$,
and represents a direct generalization of relative entropy of entanglement~\cite{PhysRevA.57.1619}.
The right side of Eq.~(\ref{eq:Pytha}) represents, respectively, the measure of
anyonic charge entanglement (ACE), $E_{\rm ACE} \equiv \widetilde{S}(\tilde{\rho} \| \Omega(\tilde{\rho}))$,
describing the entanglement arising from the non-tensor product structure of the anyonic state
spaces, and the measure of conventional entanglement (CE), $E_{\rm CE} \equiv
 \min_{\tilde{\sigma} \in {\rm SEP}} \widetilde{S} (\Omega(\tilde{\rho}) \| \tilde{\sigma})$,
arising from the usual tensor product structure of the anyonic state spaces.
Equation~(\ref{eq:Pytha}) not only provides a quantitative relationship for different types of
entanglement measures in anyonic systems, but also reveals the geometric structure of entanglement
in the anyonic state spaces since it is the application of the anyonic Pythagorean theorem
(see proof in Appendix~\ref{app:APT}).

For anyonic state $\ket{\tau, \tau; 1}$, we have
\begin{align}
    E_{\rm AREE}(\ket{\tau, \tau; 1}) & = 2 \log_2 \qd_\tau, \nonumber \\
    E_{\rm ACE} (\ket{\tau, \tau; 1}) & = 2 \log_2 \qd_\tau, \nonumber \\
    E_{\rm CE} (\ket{\tau, \tau; 1}) & =0. 
\end{align}
The $E_{\rm AREE}$ is found to be greater than the AEE,
whereas the conventional relative entropy of entanglement makes up for their difference.
According to the uniqueness theorem of entanglement~\cite{10.1063/1.1495917},
any reasonable entanglement measure should coincide with the entanglement entropy for pure states
in the asymptotic limit.
Therefore, we investigate the asymptotic behavior of these entanglement measures for
pure anyonic states.

\begin{figure}
    \includegraphics[width=1\columnwidth{}, keepaspectratio]{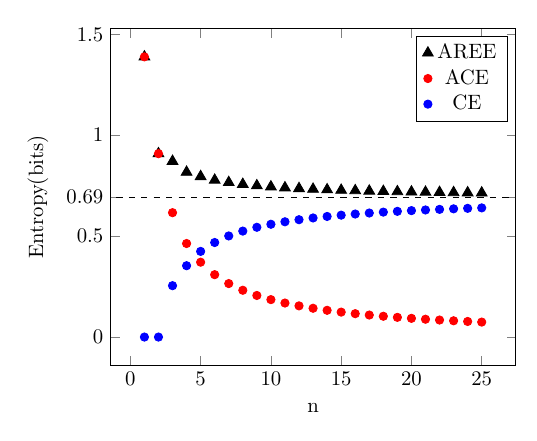}
    \caption{The scatter plots of various average entanglement measures (\ref{eq:AEM}) for
    finite $n$-copy of anyonic state $\ket{\tau, \tau; 1}$.
    Black triangles, red dots, and blue dots denote the average $E_{\rm AREE}$, $E_{\rm ACE}$,
    and $E_{\rm CE}$ respectively.
    Dashed lines indicate the vertical coordinate $y=\log_2 \qd_\tau \approx 0.69$,
    which is the asymptote for the average $E_{\rm AREE}$ and $E_{\rm CE}$.}
    \label{fig:1}
\end{figure}

We examine the related average entanglement measures:
\begin{equation}
    \label{eq:AEM}
    \frac{E_{\rm AREE}(\ket{\tau, \tau; 1}^n)}{n},~\frac{E_{\rm ACE}(\ket{\tau, \tau; 1}^n)}{n},~
    \frac{E_{\rm CE}(\ket{\tau, \tau; 1}^n)}{n},
\end{equation}
for finite $n$ copies of $\ket{\tau, \tau; 1}$.
The results are presented in Figure~\ref{fig:1}. From the figure,
we can observe that, as the number of copies increases, the $E_{\rm AREE}$ and $E_{\rm ACE}$ of
a single copy gradually decrease while the average $E_{\rm CE}$ gradually increases.
Eventually, under the asymptotic limit, the average $E_{\rm ACE}$ disappears
while the average $E_{\rm AREE}$ and $E_{\rm CE}$
converges to the AEE (the horizontal line $y = \log_2 \qd_\tau \approx 0.69$) as expected.
Notable, for the average $E_{\rm CE}$, when $n$ equals $1$ and $2$, it gives zero for both cases.

In order to confirm the aforementioned asymptotic behaviors, we obtain the following theorem.

\textbf{Theorem~2 (Asymptotic Behavior of Anyonic Entanglement)} --
For pure anyonic states $\tilde{\rho}$, the average $E_{\rm AREE}$ and $E_{\rm CE}$ approach the AEE
in the asymptotic limit, while the average $E_{\rm ACE}$ vanishes,
\begin{align}
    & \lim_{n \rightarrow \infty} \frac{E_{\rm AREE} (\tilde{\rho}^n)}{n}
        = \lim_{n \rightarrow \infty} \frac{E_{\rm CE} (\tilde{\rho}^n)}{n}
        = E_{\rm AEE}, \nonumber \\
    & \lim_{n \rightarrow \infty} \frac{E_{\rm ACE} (\tilde{\rho}^n)}{n} = 0.
\end{align}
The proof is straightforward utilizing the Schmidt decomposition (see Appendix~\ref{app:theorem2}).

These properties of entanglement measures can be intuitively understood.
On the one hand, with an increasing number of copies, the number of local operations
that can be locally performed grows even faster.
This is because the fusion rules combine these copies, relaxing not only the constraints
imposed by the anyonic superselection rules to some extent but also, due to their non-Abelian nature,
enhancing the availability of more Hilbert spaces for utilization.
On the other hand, the $E_{\rm ACE}$ represents the correlation between subsystems resulting from
the superselection rules and the non-Abelian nature of anyons~\cite{PhysRevA.108.052221}.
Therefore, in the multi-copy setting, the fusion rules cause information to flow from
the anyonic state spaces that do not satisfy the tensor product structure to those that satisfy
the tensor product structure.

Now, let us return to the initial question: How can we understand pure anyonic states
exhibiting the superactivation of nonlocality, a phenomenon previously considered unique
to conventional mixed states?

The nonlocality of anyonic states is tested using the tensor product of local observables,
which actually reflects information originating from the anyonic state spaces that satisfy the
tensor product structure.
Anyonic states $\tilde{\rho}$ and $\Omega(\tilde{\rho})$ will indeed give the same results
in the Bell test since
\begin{equation}
    \tTr \left[ \tilde{\rho} A_\alpha \otimes B_\beta \right]
        = \tTr \left[ \Omega(\tilde{\rho}) A_\alpha \otimes B_\beta \right], \nonumber
\end{equation}
for all local observables $A_\alpha$ and $B_\beta$~\cite{PhysRevA.108.052221}.
Therefore the corresponding measure of entanglement used to explain nonlocality should remain
originating from the anyonic state spaces with the tensor product structure.

The value of $E_{\rm CE}$ is a suitable candidate for measuring entanglement in this context
as it is the same for anyonic states $\tilde{\rho}$ and $\Omega(\tilde{\rho})$.
In contrast, AEE captures the total entanglement, including contributions from the anyonic spaces
that possess the non-tensor product structure, in the asymptotic limit~\cite{PhysRevA.90.062325}.
As shown in Fig.~\ref{fig:1}, when $n$ equals $1$ or $2$, the average $E_{\rm CE}$ is $0$,
which precisely explains why the corresponding anyonic states are local.
Interestingly, a non-zero AEE indicates that there exists some $n$ (here, equal to $3$)
such that the $E_{\rm CE}$ is non-zero, thereby activating the nonlocality of the corresponding
anyonic states.

\section{\label{sec:5}Proof to show that $E_{\rm CE}$ characterizes Bell nonlocality}

In this section we will demonstrate that the measure $E_{\rm CE}$ characterizes the Bell nonlocality
for pure anyonic states, as stated in the following theorem.

\textbf{Theorem~3 (Nonlocality in pure anyonic states)} --
A pure anyonic state is local iff its $E_{\rm CE} = 0$.
Furthermore, it can exhibit superactivation of nonlocality iff its $E_{\rm ACE} > 0$.

\textbf{Proof} --
We observe that in the case of pure anyonic states~(\ref{eq:SchPS}), $E_{\rm CE}$ takes the following
form (see Appendix~\ref{app:theorem2}).
\begin{equation}
    E_{\rm CE} = \sum_{c = \left\{ 1, \tau \right\}} p_c H (\left\{ \lambda_{c,i}/p_c \right\}), \nonumber 
\end{equation}
where $\left\{ \lambda_{c,i} \right\}$ are the Schmidt coefficients,
$p_c = \sum_i^{K_c} \lambda_{c,i}$ with $K_c$ denoting the Schmidt rank of sector-$c$,
and $H(\cdot)$ is the Shannon entropy.
Combining the formula of $E_{\rm ACE}$ for pure anyonic states~\cite{bonderson2017anyonic, PhysRevA.108.052221}:
\begin{equation}
    E_{\rm ACE} = H(\left\{ p_c \right\}) + \sum_c p_c \log_2 \qd_c, \nonumber
\end{equation}
we can categorize pure anyonic states into three types as follows:
\begin{align}
    {\rm type~A}:~ \ket{\rm A} = & \ket{i; 1}_A \ket{i;1}_B, \nonumber \\
    {\rm type~B}:~ \ket{\rm B} = & \sqrt{\lambda_{1,i}} \ket{i; 1}_A \ket{i;1}_B \nonumber \\
        & + \sqrt{\lambda_{\tau,j}} \ket{j; \tau}_A \ket{j; \tau}_B \ket{\tau, \tau;1}, \nonumber\\
    {\rm type~C}:~ \ket{\rm C} = & \sum_i^{K_1} \sqrt{\lambda_{1,i}} \ket{i; 1}_A \ket{i;1}_B \nonumber \\
        & + \sum_j^{K_\tau} \sqrt{\lambda_{\tau,i}} \ket{j; \tau}_A \ket{j; \tau}_B \ket{\tau, \tau;1}, \nonumber
\end{align}
where $\left\{ \ket{i; c}_A \right\}$ and $\left\{ \ket{i; \overline{c}}_B \right\}$
are the canonical bases of the Hilbert spaces of subsystems $A$ and $B$, respectively,
and $K_1 K_\tau \ge 2$.
Anyonic states of type A and type B are characterized by $E_{\rm ACE} = 0$, $E_{\rm CE}=0$
and $E_{\rm ACE}>0$, $E_{\rm CE}=0$, respectively, while anyonic states of type C satisfy  $E_{\rm CE}>0$.
It is straightforward to demonstrate that anyonic states of type A and B are local.
Next, we will establish that anyonic states of type C are nonlocal as they violate the CHSH inequality.

Since $K_1 K_\tau \ge 2$ for states of type C. We can assume that $K_1 \ge 2$ without loss of generallty.
The observables employed by Alice can be parameterized in the canonical basis
\begin{equation}
    A_\alpha = \begin{pmatrix}
        \cos \alpha & \sin \alpha & 0 & 0 & \cdots \\
        \sin \alpha & -\cos \alpha & 0 & 0 & \cdots \\
        0 & 0 & 1 & 0 & \cdots \\
        0 & 0 & 0 & 1 & \cdots \\
        \vdots & \vdots & \vdots & \vdots & \ddots
    \end{pmatrix}, \nonumber
\end{equation}
where the basis vectors corresponding to the non-zero elements in the first two rows of $A_\alpha$
are $\ket{1;1}_A$ and $\ket{2;1}_A$, respectively.
Bob's observables $B_\beta$ are defined in the same way.
Thus, we have
\begin{align}
    & \langle A_{\alpha_1} B_{\beta_1} \rangle + \langle A_{\alpha_1} B_{\beta_2} \rangle
    + \langle A_{\alpha_2} B_{\beta_1} \rangle - \langle A_{\alpha_2} B_{\beta_2} \rangle \nonumber \\
    = & 2p\sqrt{1+ \frac{4\lambda_{1,1}\lambda_{1,2}}{p^2}}
        + 2(1-p) > 2,
\end{align}
where $p = \lambda_{1,1}+\lambda_{1,2}$, $\alpha_1 = 0$, $\alpha_2 = \frac{\pi}{2}$,
and $\beta_1 = - \beta_2 = {\rm tan}^{-1}\frac{\lambda_{1,1}+\lambda_{1,2}}{2\sqrt{\lambda_{1,1}\lambda_{1,2}}}$.

Since anyonic states of type B have non-zero $E_{\rm ACE}$, the corresponding multi-copy states attain
non-zero $E_{\rm CE}$ that converges to AEE (Theorem~2), thereby violating the CHSH inequality.
Hence, local anyonic states with non-zero $E_{\rm ACE}$ exhibit superactivation of nonlocality,
which completes the proof of Theorem 3.

\section{Summary}

In this paper, we systematically investigate the quantum nonlocality of pure anyonic states
and discover the analogous phenomenon of superactivation of nonlocality.
Specifically, we provide an operational scheme for testing Bell inequalities using multiple
copies of anyonic states.
By examining the anyonic state $\ket{\tau, \tau; 1}$, which has non-zero AEE,
we find that it is local when the number of copies is less than $3$.
However, when the number of copies exceeds $3$, the state exhibits nonlocality.
For conventional quantum information theory, this is not a trivial phenomenon
because the necessary and sufficient condition for conventional pure quantum states to
possess nonlocality is that they have non-zero entanglement entropy,
which is a characteristic feature of conventional mixed states.

To explain this non-trivial phenomenon, we investigated the asymptotic behavior of
anyonic entanglement.
By utilizing the decomposition formula of anyonic entanglement (\ref{eq:Pytha}),
we find that a portion of the entanglement $E_{\rm ACE}$ stemming from the non-tensor product
structure of the anyonic state spaces gradually transforms into conventional
entanglement $E_{\rm CE}$ (originating from the tensor product structure of the anyonic state spaces),
which converges to the AEE in the asymptotic limit.
The superactivation of nonlocality corresponds to the abrupt transition
where the average $E_{\rm CE}$ seems to emerge from nothing.
A nonzero AEE indicates the existence of such a transition.
We can thus readily conclude that any anyonic pure state, apart from product states,
must either show nonlocality or exhibit superactivation of nonlocality—phenomena identified by
the nonzero values for the measures $E_{\rm CE}$ and $E_{\rm ACE}$, respectively.

Our work demonstrates that anyonic systems, whose state spaces lack a tensor product
structure, possess entanglement with a more intricate structure.
We hope our work will shed new light on anyonic quantum information.

\begin{acknowledgments}
   C.Q. Xu is grateful to Yilong Wang for helpful discussions.
   This work is supported by the Innovation Program for Quantum Science and
   Technology (2021ZD0302100) and by NSFC (Grants No. 12361131576 and No. 92265205).
   C.Q. Xu is also supported by NSFC (Grant No. 12405018)
   and by the Postdoctoral Fellowship Program of CPSF (Grant No. GZC20231366).
\end{acknowledgments}

\appendix

\section{\label{app:Sym} The Entanglement Symmetry of Pure Anyonic States}

In this section we will show that the entanglement symmetry of pure anyonic states
with Abelian total charge.

With the help of Schmidt decomposition, an pure anyonic state with Abelian total charge $a$
can be written as
\begin{equation}
    \ket{\tilde{\psi}}_{AB} = \sum_{a, X, Y, \mu} \sum_{i=1}^{K_{aXY\mu}} \sqrt{\lambda_{aXYi\mu}} 
        \ket{(i; X)_A, (i; Y)_B ; a, \mu}, \nonumber
\end{equation}
where the subscripts $A$ and $B$ indicate the subsystems $A$ and $B$,
and $\lambda_{aXYi\mu}$ are non-negative real numbers called Schmidt coefficients
that satisfy the normalization condition $\sum_{a, X, Y, i,\mu} \lambda_{aXYi\mu} = 1$.
Charges $a$, $X$ and $Y$ are restricted by fusion rule
\begin{equation}
    X \times Y = N_{XY}^a a + \cdots, 
\end{equation}
where fusion coefficient $N_{XY}^a \ge 1$ and ellipses indicate that there may be other
fusion results.
In the following we will show that, given an Abelian charge $a$ and a charge $X$,
there is only one charge $Y$ such that the fusion result of $X$ and $Y$ has charge $a$
with fusion coefficient $N_{XY}^a = 1$,
and the quantum dimensions of $X$ and $Y$ are equal.

Using the diagrammatic representation, we can represent this process by which
charges $X$ and $Y$ fuse to get Abelian charge $a$:
\begin{equation}
    \begin{tikzpicture}[baseline]
        \draw (-0.5, -0.5) node[below]{$X$} -- (0,0)  node[sloped, pos = 0.5]{\arrowIn}
            -- (0.5,-0.5) node[below]{$Y$}  node[sloped, pos = 0.5]{\arrowOut};
        \draw (0,0) node[right]{$\mu$} -- (0,0.5) node[above]{$a$} node[sloped, pos=0.5]{\arrowIn};
    \end{tikzpicture},
\end{equation}
where $\mu = 1, \cdots, N_{XY}^a$.
We can bend the downturned line $X$ into the upturned line $\overline{X}$ using the pivotal
property of anyon model:
\begin{equation}
    \begin{tikzpicture}[baseline]
        \draw (-1,-0.293) arc(180:315:0.293);
        \draw (-1,0.5) node[above]{$\overline{X}$} -- (-1,-0.293) node[sloped, pos=0.5]{\arrowIn};
        \draw (-0.5, -0.5) node[below]{$X$} -- (0,0)  node[sloped, pos = 0.5]{\arrowIn}
            -- (0.5,-0.5) node[below]{$Y$}  node[sloped, pos = 0.5]{\arrowOut};
        \draw (0,0) node[right]{$\mu$} -- (0,0.5) node[above]{$a$} node[sloped, pos=0.5]{\arrowIn};
    \end{tikzpicture} = \sum_{\nu} \left[ U_{XY}^a \right]_{\mu \nu}
    \begin{tikzpicture}[baseline]
        \draw (-0.5, 0.5) node[above]{$\overline{X}$} -- (0,0)  node[sloped, pos = 0.5]{\arrowIn}
            -- (0.5,0.5) node[above]{$a$}  node[sloped, pos = 0.5]{\arrowIn};
        \draw (0,0) node[right]{$\nu$} -- (0,-0.5) node[below]{$Y$} node[sloped, pos=0.5]{\arrowOut};
    \end{tikzpicture},
\end{equation}
where $U_{XY}^a$ is unitary in indexes $\mu$ and $\nu$.
The process on the right side of the above equation corresponds to the fusion rule
\begin{equation}
    \overline{X} \times a = N_{\overline{X}a}^Y Y + \cdots.
\end{equation}
It can be inferred that the fusion coefficient $N_{\overline{X}a}^Y =N_{XY}^a= 1$
and there are no other charge can be fused except charge $a$ since charge $a$ is Abelian.
In other words, given an Abelian charge $a$ and a charge $X$,
there is only one charge $Y$ such that the fusion result of $X$ and $Y$ has charge $a$,
and equality for quantum dimensions
\begin{equation}
    \qd_{\overline{X}} \qd_a = \qd_Y = \qd_X.
\end{equation}

Now the Schmidt decomposition of pure anyonic state can be simplified as
\begin{equation}
    \ket{\tilde{\psi}}_{AB} = \sum_{a, X} \sum_{i=1}^{K_{aX}} \sqrt{\lambda_{aXi}} 
        \ket{(i; X)_A, (i; Y)_B; a}, \nonumber
\end{equation}
where $Y = \overline{X} \times a$.
By tracing out the degrees of freedom of subsystem $B$ ($A$), we can obtain
the anyonic state of subsystem $A$ ($B$):
\begin{align}
    \tilde{\rho}_A & = \sum_X \sum_{i=1}^{K_{aX}} \frac{\lambda_{aXi}}{\qd_X} \ket{i; X}_A \bra{i; X}, \nonumber \\
    \tilde{\rho}_B & = \sum_X \sum_{i=1}^{K_{aX}} \frac{\lambda_{aXi}}{\qd_Y} \ket{i; Y}_B \bra{i; Y},
\end{align}
which have same anyonic entropy
\begin{align}
    & \widetilde{S}(\tilde{\rho}_A) = \widetilde{S}(\tilde{\rho}_B) \nonumber \\
    = & - \sum_{a, X, i} \lambda_{aXi} \log_2 \lambda_{aXi}
    + \sum_{X} p_X \log_2 \qd_X,
\end{align}
where $p_X = \sum_{a, i} \lambda_{aXi}$.

\section{\label{app:TB} The Tangled Braiding and Multiple Copies of an Anyoinc State}

In this section, we provide more precise definitions including multiple copies of
a bipartite anyonic state and the tangled braiding $\mathcal{T}$.
and the proofs of equalities in the main text.
We restrict ourselves to the anyon models on a sphere with fully isotopy.

In the following, we present two definitions of multiple copies of an anyonic state.

\textbf{Definition~B1 (Canonical Multiple Copies)} --
For a bipartite anyonic state $\tilde{\rho}_{AB}$, we define canonical $n$-copy
of $\tilde{\rho}_{AB}$ as
\begin{equation}
    \tilde{\rho}_{AB}^{\otimes n} =
        \begin{tikzpicture}[baseline, scale=0.5]
            \draw (-2.5,0.5) rectangle (-0.5,-0.5); \draw (0.5,0.5) rectangle (2.5,-0.5);
            \draw (0,0) node[]{$\otimes$}; \draw (3.5,0) node[]{$\otimes \cdots$};
            \draw (-1.5,0) node[]{$AB$}; \draw (1.5,0) node[]{$AB$};
            \draw (-0.5,0.5) -- (-0.5,1) node[above]{$B_1$}; \draw (-2.5,0.5) -- (-2.5,1) node[above]{$A_1$};
            \draw (-0.5,-0.5) -- (-0.5,-1) node[below]{$B_1$}; \draw (-2.5,-0.5) -- (-2.5,-1) node[below]{$A_1$};
            \draw (0.5,0.5) -- (0.5,1) node[above]{$A_2$}; \draw (2.5,0.5) -- (2.5,1) node[above]{$B_2$};
            \draw (0.5,-0.5) -- (0.5,-1) node[below]{$A_2$}; \draw (2.5,-0.5) -- (2.5,-1) node[below]{$B_2$};
        \end{tikzpicture},
\end{equation}
where we have used rectangles to represent the anyonic density matries $\tilde{\rho}_{AB}$
and outer legs to represent the corresponding subsystems composed of anyons.

\textbf{Definition~B2 (Joint Multiple Copies)} --
We define joint $n$-copy of $\tilde{\rho}_{AB}$ as
\begin{equation}
    \tilde{\rho}_{AB}^n =
        \begin{tikzpicture}[baseline, scale=0.5]
            \draw (-1.2,0.5) rectangle (1.2,-0.5); \draw (0,0) node[]{$AB$};
            \draw (-0.8,1.3) rectangle (0.8,0.5); \draw (0,0.9) node[]{$AB$};
            \draw (-0.8,-0.5) rectangle (0.8,-1.3);
            \draw (-1.2,0.5) -- (-2,2) node[above]{$A_1$};
            \draw (1.2,0.5) -- (2,2) node[above]{$B_1$};
            \draw (-0.8,1.3) -- (-1.2,2) node[above]{$A_2$};
            \draw (0.8,1.3) -- (1.2,2) node[above]{$B_2$};
            \draw (-1.2,-0.5) -- (-2,-2) node[below]{$A_1$};
            \draw (1.2,-0.5) -- (2,-2) node[below]{$B_1$};
            \draw (-0.8,-1.3) -- (-1.2,-2) node[below]{$A_2$};
            \draw (0.8,-1.3) -- (1.2,-2) node[below]{$B_2$};
            \draw (0,1.8) node[]{$\cdots$};
        \end{tikzpicture}.
\end{equation}

We can think of the $\tilde{\rho}_{AB}^{\otimes n}$ as the side-by-side placement of
anyonic state $\tilde{\rho}_{AB}$, while the $\tilde{\rho}_{AB}^n$ is the front-and-back
placement of $\tilde{\rho}_{AB}$.
These two states $\tilde{\rho}_{AB}^{\otimes n}$ and $\tilde{\rho}_{AB}^n$ should
describe the same state of two-dimensional system even though they have different bases.
Here, we give the corresponding transformation between these two states.
To this end, let's define the tangled braidng $\mathcal{T}$~\cite{PhysRevA.106.012413}.

\textbf{Definition~B3 (Tangled Braiding)} --
The tangled braiding $\mathcal{T}$ is a series of
braidings $\mathcal{T} = T_n T_{n-1} \cdots T_2$,
where braiding $T_k$ ($k \in \{ 2,3,\cdots,n \}$) is defined as
\begin{align}
  \label{eq:Tk}
  T_k \equiv
  \begin{tikzpicture}[baseline, scale = 0.6]
    \draw (-1,1) -- (-1,-1);
    \draw (-1.5,1) -- (-1.5,-1);
    \draw (-2.5,1) -- (-2.5,-1);
    \draw[dotted] (1.8,1)--(2.2,1);
    \draw[dotted] (1.3,-1)--(1.7,-1);
    \draw[dotted] (-1.8,1)--(-2.2,1);
    \draw[dotted] (-1.8,-1)--(-2.2,-1);
    \draw (2,-1)--(2.5,1);
    \draw (1,-1)--(1.5,1);
    \draw (0.5,-1)--(1,1);
    \draw (3.5,-1)--(2.3,-0.2); \draw (2.1,-0.07) -- (1.5,0.33);
    \draw (1.25,0.5) -- (1,0.67); \draw (0.9,0.73) -- (0.5,1);
    \draw (2.5,-1) -- (2.15,-0.77); \draw (2,-0.67) -- (1.3,-0.2);
    \draw (1.1,-0.07) -- (0.9,0.07); \draw (0.7,0.2) -- (-0.5,1);
    \draw [decorate,decoration={brace,amplitude=10pt},xshift=0pt,yshift=6pt]
    (1,1)-- node[midway,yshift=0.6cm]{$k-1$}(2.5,1);
    \draw [decorate,decoration={brace,amplitude=10pt},xshift=0pt,yshift=-6pt]
    (-1,-1)-- node[midway,yshift=-0.6cm]{$k-1$}(-2.5,-1);
  \end{tikzpicture},
\end{align}
where the solid line can represent either a single anyon or a subsystem composed of anyons.

The tangled braiding $\mathcal{T}$ can transform the canonical $n$-copy into
the joint $n$-copy of $\tilde{\rho}_{AB}$:
\begin{equation}
    \label{eq:B4}
    \tilde{\rho}_{AB}^n = \mathcal{T} \tilde{\rho}_{AB}^{\otimes n} \mathcal{T}^\dagger.
\end{equation}
Here we prove Eq.~(\ref{eq:B4}) for $n=2$,
\begin{align}
    & \mathcal{T} \tilde{\rho}_{AB}^{\otimes 2} \mathcal{T}^\dagger \nonumber \\
    = & \begin{tikzpicture}[baseline, scale=0.5]
            \draw (-2.5,0.5) rectangle (-0.5,-0.5); \draw (0.5,0.5) rectangle (2.5,-0.5);
            \draw (-1.5,0) node[]{$AB$}; \draw (1.5,0) node[]{$AB$};
            \draw (-0.5,0.5) -- (-0.5,1) -- (2.5,3) node[above]{$B_1$};
            \draw (-2.5,0.5) -- (-2.5,3) node[above]{$A_1$};
            \draw (-0.5,-0.5) -- (-0.5,-1) -- (2.5,-3) node[below]{$B_1$};
            \draw (-2.5,-0.5) -- (-2.5,-3) node[below]{$A_1$};
            \draw (0.5,0.5) -- (0.5,1) -- (0.2,1.3); \draw (2.5,0.5) -- (2.5,1) -- (1.4,2.1);
            \draw (0,1.5) -- (-1.5,3) node[above]{$A_2$}; \draw (1.2,2.3) -- (0.5,3) node[above]{$B_2$};
            \draw (0.5,-0.5) -- (0.5,-1) -- (0.2,-1.3); \draw (2.5,-0.5) -- (2.5,-1) -- (1.4,-2.1);
            \draw (0,-1.5) -- (-1.5,-3) node[below]{$A_2$};
            \draw (1.2,-2.3) -- (0.5,-3) node[below]{$B_2$};
        \end{tikzpicture} =
        \begin{tikzpicture}[baseline, scale=0.5]
            \draw (-1.2,0.5) rectangle (1.2,-0.5); \draw (0,0) node[]{$AB$};
            \draw (-0.8,1.3) rectangle (0.8,0.5); \draw (0,0.9) node[]{$AB$};
            \draw (-0.8,-0.5) rectangle (0.8,-1.3);
            \draw (-1.2,0.5) -- (-2,2) node[above]{$A_1$};
            \draw (1.2,0.5) -- (2,2) node[above]{$B_1$};
            \draw (-0.8,1.3) -- (-1.2,2) node[above]{$A_2$};
            \draw (0.8,1.3) -- (1.2,2) node[above]{$B_2$};
            \draw (-1.2,-0.5) -- (-2,-2) node[below]{$A_1$};
            \draw (1.2,-0.5) -- (2,-2) node[below]{$B_1$};
            \draw (-0.8,-1.3) -- (-1.2,-2) node[below]{$A_2$};
            \draw (0.8,-1.3) -- (1.2,-2) node[below]{$B_2$};
            \end{tikzpicture} \nonumber \\
    = & \tilde{\rho}_{AB}^2.
\end{align}
which can be generalized to other cases
using recursion $T_k (\tilde{\rho}_{AB}^{k-1} \otimes \tilde{\rho}_{AB}) T_k^\dagger = \tilde{\rho}_{AB}^k$.

The tangled braiding $\mathcal{T}$ ensures that Alice's trace-preserving
local operations $\mathcal{E}_A$ will not affect Bob's subsystems:
\begin{equation}
    \label{eq:B6}
    \widetilde{\operatorname{Tr}}_A \left[ \mathcal{E}_A(\mathcal{T} \tilde{\rho}_{AB}^{\otimes n}
        \mathcal{T}^\dagger) \right] = \tilde{\rho}_B^{\otimes n}.
\end{equation}
Here we prove Eq.~(\ref{eq:B6}) for $n=2$:
\begin{widetext}
    \begin{align}
        \widetilde{\operatorname{Tr}}_A \left[ \mathcal{E}_A(\mathcal{T} \tilde{\rho}_{AB}^{\otimes n}
            \mathcal{T}^\dagger) \right]
        = \sum_i \begin{tikzpicture}[baseline, scale=0.5]
            \draw (-2.5,0.5) rectangle (-0.5,-0.5); \draw (0.5,0.5) rectangle (2.5,-0.5);
            \draw (-1.5,0) node[]{$AB$}; \draw (1.5,0) node[]{$AB$};
            \draw (-0.5,0.5) -- (-0.5,1) -- (2.5,3) -- (2.5,4); \draw (-2.5,0.5) -- (-2.5,3);
            \draw (-0.5,-0.5) -- (-0.5,-1) -- (2.5,-3) -- (2.5,-4); \draw (-2.5,-0.5) -- (-2.5,-3);
            \draw (0.5,0.5) -- (0.5,1) -- (0.2,1.3); \draw (2.5,0.5) -- (2.5,1) -- (1.4,2.1);
            \draw (0,1.5) -- (-1.5,3); \draw (1.2,2.3) -- (0.5,3) -- (0.5,4);
            \draw (0.5,-0.5) -- (0.5,-1) -- (0.2,-1.3); \draw (2.5,-0.5) -- (2.5,-1) -- (1.4,-2.1);
            \draw (0,-1.5) -- (-1.5,-3); \draw (1.2,-2.3) -- (0.5,-3) -- (0.5,-4);
            \draw (-2.7,3) rectangle (-1.3,4); \draw (-2,3.5) node[]{$E_i$};
            \draw (-2.7,-3) rectangle (-1.3,-4); \draw (-2,-3.5) node[]{$E_i^\dagger$};
            \draw (-2.5,4) -- (-3,4.5) -- (-3.5,4) -- (-3.5,-4) -- (-3,-4.5) -- (-2.5,-4);
            \draw  (-1.5,4) -- (-3,5.5) -- (-4.5,4) -- (-4.5,-4) -- (-3,-5.5) -- (-1.5,-4);
        \end{tikzpicture} =
        \begin{tikzpicture}[baseline, scale=0.5]
            \draw (-2.5,0.5) rectangle (-0.5,-0.5); \draw (0.5,0.5) rectangle (2.5,-0.5);
            \draw (-1.5,0) node[]{$AB$}; \draw (1.5,0) node[]{$AB$};
            \draw (-0.5,0.5) -- (-0.5,1) -- (2.5,3) -- (2.5,4); \draw (-2.5,0.5) -- (-2.5,3);
            \draw (-0.5,-0.5) -- (-0.5,-1) -- (2.5,-3) -- (2.5,-4); \draw (-2.5,-0.5) -- (-2.5,-3);
            \draw (0.5,0.5) -- (0.5,1) -- (0.2,1.3); \draw (2.5,0.5) -- (2.5,1) -- (1.4,2.1);
            \draw (0,1.5) -- (-1.5,3); \draw (1.2,2.3) -- (0.5,3) -- (0.5,4);
            \draw (0.5,-0.5) -- (0.5,-1) -- (0.2,-1.3); \draw (2.5,-0.5) -- (2.5,-1) -- (1.4,-2.1);
            \draw (0,-1.5) -- (-1.5,-3); \draw (1.2,-2.3) -- (0.5,-3) -- (0.5,-4);
            \draw (-2.5,3) -- (-2.5,4) -- (-3,4.5) -- (-3.5,4) -- (-3.5,-4)
                    -- (-3,-4.5) -- (-2.5,-4) -- (-2.5,-3);
            \draw  (-1.5,3) -- (-1.5,4) -- (-3,5.5) -- (-4.5,4) -- (-4.5,-4)
                    -- (-3,-5.5) -- (-1.5,-4) -- (-1.5,-3);
        \end{tikzpicture} =
        \begin{tikzpicture}[baseline, scale=0.5]
            \draw (-2.5,0.5) rectangle (-0.5,-0.5); \draw (0.5,0.5) rectangle (2.5,-0.5);
            \draw (-1.5,0) node[]{$B$}; \draw (1.5,0) node[]{$B$};
            \draw (-1.5,0.5) -- (-1.5,3); \draw (-1.5,-0.5) -- (-1.5,-3);
            \draw (1.5,0.5) -- (1.5,3); \draw (1.5,-0.5) -- (1.5,-3);
        \end{tikzpicture} = \tilde{\rho}_B^{\otimes 2},
    \end{align}
\end{widetext}
where we have used Kraus operators $\left\{ E_i \right\}$ to denote quantum
operations $\mathcal{E}_A(\cdot) = \sum_i E_i (\cdot) E_i^\dagger$
and used the completeness relation $\sum_i E_i^\dagger E_i = 1$
of trace-preserving quantum operations.

Using Eq.~(\ref{eq:B6}), we can obtain the AEE of
state $\mathcal{T} \tilde{\rho}_{AB}^{\otimes n} \mathcal{T}^\dagger$:
\begin{equation}
    E_{\rm AEE}(\mathcal{T} \tilde{\rho}_{AB}^{\otimes n} \mathcal{T}^\dagger)
    = \widetilde{S}(\tilde{\rho}_A^{\otimes n}) = n E_{\rm AEE}(\tilde{\rho}_{AB}),
\end{equation}
where $\widetilde{S}(\tilde{\sigma}) = - \tTr \left(
 \tilde{\sigma}\log_2 \tilde{\sigma} \right)$ is anyonic entropy of anyonic
state $\tilde{\sigma}$, and AEE of anyonic state $\tilde{\sigma}_{AB}$ can be
defined as $E_{\rm AEE} = \widetilde{S}(\tilde{\sigma}_A)$.
The above result is intuitive: the entanglement entropy of a $n$-copied system 
is $n$ times the entanglement entropy of a single.

Here we do not distinguish whether $\tilde{\rho}_{AB}$ is pure or mixed.
When $\tilde{\rho}_{AB}$ is mixed, however, it is crucial to consider the location
of the external anyons correlated with $\tilde{\rho}_{AB}$ before manipulation.
Using the properties of the tangled braiding $\mathcal{T}$, we provide an
operational scheme as follows to circumvent this problem.
\begin{itemize}
    \item \textbf{Purification}: We treat the external anyons as subsystem $C$,
           which together with $\tilde{\rho}_{AB}$ becomes a pure state $\ket{ABC}$.
    \item \textbf{Preparation}: We independently prepare multiple copies
          of $\ket{ABC}$: $\ket{ABC}^{\otimes n}$.
    \item \textbf{Braiding}: We apply the tangled braiding $\mathcal{T}$
          to separate the multiple copies of subsystems $AB$ and $C$.
\end{itemize}
Finally we can get $\tilde{\rho}_{AB}^n$ and $\tilde{\rho}_C^n$.

\section{\label{app:APT} Anyonic Pythagorean Theorem for Anyonic Relative Entropy}

In this section, we will prove the following theorem based on the anyonic relative
entropy between anyonic states $\tilde{\rho}$ and $\tilde{\sigma}$:
 $\widetilde{S}(\tilde{\rho} \parallel \tilde{\sigma}) = \tTr \left( \tilde{\rho}
 \log_2 \tilde{\rho} \right) - \tTr \left( \tilde{\rho} \log_2 \tilde{\sigma} \right)$,
and show that Eq.~(\ref{eq:Pytha}) in the main text is an application of this theorem.

\textbf{Theorem~C1 (Anyonic Pythagorean Theorem for Anyonic Relative Entropy)} --
For three anyonic states $\tilde{\rho}_1$, $\tilde{\rho}_2$ and $\tilde{\rho}_3$,
satisfying ${\rm supp}(\tilde{\rho}_1) \subseteq {\rm supp}(\tilde{\rho}_2)
 \subseteq {\rm supp}(\tilde{\rho}_3)$, we have the following relation:
\begin{align}
    \label{eq:C1}
    & \widetilde{S}(\tilde{\rho}_1 \parallel \tilde{\rho}_3)
        - \widetilde{S}(\tilde{\rho}_1 \parallel \tilde{\rho}_2)
        -\widetilde{S}(\tilde{\rho}_2 \parallel \tilde{\rho}_3) \nonumber \\
    = & \sum_\alpha (\eta_1^\alpha - \eta_2^\alpha)(\theta_1^\alpha - \theta_3^\alpha),
\end{align}
where real parameters $\theta$ and $\eta$ are derived from two forms of states,
respectively,
\begin{align}
    \label{eq:stateform}
    & \tilde{\rho}_{\rm e} = e^{\sum_\alpha \theta^\alpha \sigma_\alpha}, \\
    & \tilde{\rho}_{\rm a} = \sum_\alpha \eta^\alpha \sigma_\alpha.
\end{align}
where, for $\alpha \neq 0$, $\sigma_\alpha$ are Hermitian, traceless and mutually
orthogonal operators, and $\sigma_0$ is the identity operator.

Theorem~C1 is the anyonic version of Pythagorean theorem for relative
entropy [see Eq.~(10) in Ref.~\cite{PhysRevA.80.022113}].

\textbf{Proof} --
Substituting the above two forms of the anyonic state into the anyonic relative
entropy between $\tilde{\rho}_1$ and $\tilde{\rho}_2$, we obtain
\begin{equation}
    \widetilde{S} (\tilde{\rho}_1 \parallel \tilde{\rho}_2) =
        \sum_\alpha \eta_1^\alpha \theta_1^\alpha - \sum_\alpha \eta_1^\alpha \theta_2^\alpha,
\end{equation}
where we have used the orthogonal condition for operators $\sigma$:
 $\tTr \left( \sigma_\alpha^\dagger \sigma_{\alpha'} \right) = \delta_{\alpha\alpha'}$.
By direct calculattion, we have
\begin{align}
    & \widetilde{S}(\tilde{\rho}_1 \parallel \tilde{\rho}_3)
        - \widetilde{S}(\tilde{\rho}_1 \parallel \tilde{\rho}_2)
        -\widetilde{S}(\tilde{\rho}_2 \parallel \tilde{\rho}_3) \nonumber \\
    = & - \sum_\alpha \eta_2^\alpha \theta_2^\alpha - \sum_\alpha \eta_1^\alpha \theta_3^\alpha
        + \sum_\alpha \eta_1^\alpha \theta_2^\alpha + \sum_\alpha \eta_2^\alpha \theta_3^\alpha
        \nonumber \\
    = & \sum_\alpha (\eta_1^\alpha - \eta_2^\alpha) (\theta_2^\alpha - \theta_3^\alpha).
\end{align}

The equality
\begin{equation}
    \min_{\tilde{\sigma} \in {\rm SEP}} \widetilde{S} (\tilde{\rho} \parallel \tilde{\sigma})
        = \widetilde{S}(\tilde{\rho} \parallel \Omega(\tilde{\rho})) + \min_{\tilde{\sigma} \in {\rm SEP}}
        \widetilde{S} (\Omega(\tilde{\rho}) \parallel \tilde{\sigma}),
\end{equation}
in the main text, is an application of Theorem~C1, which can be seen from the following
analysis.

The operators $\left\{ \sigma_\alpha \right\}$ can be divided into two sets.
The first set $\mathcal{C}_1$ contains operators formed by the tensor prodcut of
operators (observables) from each of two subsystems, such as:
\begin{equation}
    \sigma_\alpha^A \otimes \sigma_\beta^B,
\end{equation}
where $A$ ($B$) denotes the subsystem $A$ ($B$).
The second set $\mathcal{C}_2$ contains the remaining operators that cannot be represented
by tensor product.
It is noted that states $\tilde{\rho}$ and $\Omega(\tilde{\rho})$ have the same expected
value for the oeprators in set $\mathcal{C}_1$:
\begin{equation}
    \tTr \left( \tilde{\rho} \sigma_\alpha^A \otimes \sigma_\beta^B \right)
        = \tTr \left[ \Omega(\tilde{\rho}) \sigma_\alpha^A \otimes \sigma_\beta^B \right],
\end{equation}
and the fact that states $\Omega(\tilde{\rho})$ and $\tilde{\sigma} \in {\rm SEP}$ do not
have the structure of the operators in set $\mathcal{C}_2$.
We can easily verify that the right-hand side of Eq.~(\ref{eq:C1}) vanishes.

Finally, we need to specify that the operators in set $\mathcal{C}_2$ satisfy the properties
of being Hermitian, traceless and orthogonal.
We can construct the following pair of operators up to a constant:
\begin{align}
    & \tilde{\sigma} = \begin{tikzpicture}[baseline]
        \draw (-0.75, 0.5) node[above]{$a$} -- (-0.25,0) -- (-0.75,-0.5) node[below]{$a'$};
        \draw (-0.25,0) -- (0.25,0) node[pos=0.5, above]{$c$};
        \draw (0.75, 0.5) node[above]{$b$} -- (0.25,0) -- (0.75,-0.5) node[below]{$b'$};
    \end{tikzpicture} + \begin{tikzpicture}[baseline]
        \draw (-0.75, 0.5) node[above]{$a'$} -- (-0.25,0) -- (-0.75,-0.5) node[below]{$a$};
        \draw (-0.25,0) -- (0.25,0) node[pos=0.5, above]{$c$};
        \draw (0.75, 0.5) node[above]{$b'$} -- (0.25,0) -- (0.75,-0.5) node[below]{$b$};
    \end{tikzpicture}, \nonumber \\
    & \tilde{\sigma}^* =  i \begin{tikzpicture}[baseline]
        \draw (-0.75, 0.5) node[above]{$a$} -- (-0.25,0) -- (-0.75,-0.5) node[below]{$a'$};
        \draw (-0.25,0) -- (0.25,0) node[pos=0.5, above]{$c$};
        \draw (0.75, 0.5) node[above]{$b$} -- (0.25,0) -- (0.75,-0.5) node[below]{$b'$};
    \end{tikzpicture} - i \begin{tikzpicture}[baseline]
        \draw (-0.75, 0.5) node[above]{$a'$} -- (-0.25,0) -- (-0.75,-0.5) node[below]{$a$};
        \draw (-0.25,0) -- (0.25,0) node[pos=0.5, above]{$c$};
        \draw (0.75, 0.5) node[above]{$b'$} -- (0.25,0) -- (0.75,-0.5) node[below]{$b$};
    \end{tikzpicture},
\end{align}
where we have omitted the arrows as well as the variables within subsystems.
We can easily verify that the pairs of operators constructed in this way satisfy
the aforementioned properties.

\section{\label{app:theorem2}Proof of Theorem~2: Asymptotic Behavior of Anyonic Entanglement}

In this section we will prove the following theorem (Theorem~2 in the main text).

\textbf{Theorem~D1 (Asymptotic Behavior of Anyonic Entanglement)} --
For pure anyonic states $\tilde{\rho}$, the average $E_{\rm AREE}$ and $E_{\rm CE}$ approach the AEE
in the asymptotic limit, while the average $E_{\rm ACE}$ vanishes,
\begin{align}
    & \lim_{n \rightarrow \infty} \frac{E_{\rm AREE} (\tilde{\rho}^n)}{n}
        = \lim_{n \rightarrow \infty} \frac{E_{\rm CE} (\tilde{\rho}^n)}{n}
        = E_{\rm AEE}, \nonumber \\
    & \lim_{n \rightarrow \infty} \frac{E_{\rm ACE} (\tilde{\rho}^n)}{n} = 0. \nonumber
\end{align}

Before proving, we first calculate the $E_{\rm AREE}$ for the pure anyonic state, which is the minimum of
the anyonic relative entropy among the separable anyonic states.
For bipartite pure anyonic state,
\begin{align}
    \label{eq:D1}
    \ket{\tilde{\psi}} = & \sum_i \sqrt{\lambda_{1i}} \ket{i;1}_A \ket{i;1}_B \nonumber \\
    & + \sum_j \sqrt{\lambda_{\tau j}} \ket{j; \tau}_A \ket{j; \tau}_B \ket{\tau, \tau; 1},
\end{align}
we can construct the corresponding separable state
\begin{align}
    \label{eq:D2}
    \tilde{\rho}_0 = & \sum_i \lambda_{1i} \ket{i;1}_A \bra{i; 1} \otimes \ket{i;1}_B \bra{i; 1} \nonumber \\
    & + \sum_j \lambda_{\tau j} \frac{1}{\qd_\tau} \ket{j;\tau}_A \bra{j; \tau} \otimes
        \frac{1}{\qd_\tau} \ket{j;\tau}_B \bra{j; \tau}.
\end{align}
And we claim the following lemma.

\textbf{Lemma~D2} --
The $E_{\rm AREE}$ for the state $\ket{\tilde{\psi}}$ is the anyonic relative entropy
between $\tilde{\rho} = \ket{\tilde{\psi}}\bra{\tilde{\psi}}$ and $\tilde{\rho}_0$:
\begin{align}
    \widetilde{S}(\tilde{\rho} \parallel \tilde{\rho}_0) = &
        - \tTr \left( \tilde{\rho} \log_2 \tilde{\rho}_0 \right) \nonumber \\
    = & \sum_i - \lambda_{1i} \log_2 \lambda_{1i} + \sum_j - \lambda_{\tau j} \log_2 \lambda_{\tau j} \nonumber \\
    & + 2 \sum_j \lambda_{\tau j} \log_2 \qd_\tau. \nonumber
\end{align}
Next, we will prove Lemma~D2 that $\tilde{\rho}_0$ is the separable anyonic state that minimizes
the anyonic relative entropy.
The proof method is essentially the same as that used in Ref.~\cite{PhysRevA.57.1619}.
Specifically, we will prove that the gradient around the anyonic state $\tilde{\rho}_0$
\begin{equation}
    \partial_x \widetilde{S}\left( \tilde{\rho} \parallel (1-x) \tilde{\rho}_0 + x \tilde{\rho}' \right) |_{x=0},
\end{equation}
where $\tilde{\rho}' \in {\rm SEP}$, is non-negative.
In order to highlight the characteristics of anyonic systems and for the sake of brevity,
we here only investigate the part of the subsystem in the pure state that has a charge $\tau$,
\begin{equation}
    \ket{\tilde{\psi}} = \sum_j \sqrt{\lambda_{\tau j}} \ket{j; \tau}_A \ket{j; \tau}_B \ket{\tau, \tau; 1}, 
\end{equation}
because the part of the subsystem in the pure state that has a vacuum charge is the same
as the conventional pure state.

\textbf{Proof (Lemma~D2)} --
By using the representation for any positive
operator $A$, $\log A = \int_0^\infty (At -1 )/(A+t) dt/(1+t^2)$,
and letting $f(x, \tilde{\rho})
 = \widetilde{S}\left( \tilde{\rho} \parallel \tilde{\rho}_x \right)$,
where $ \tilde{\rho}_x = (1-x) \tilde{\rho}_0 + x \tilde{\rho}'$, we have
\begin{align}
    \partial_x f(x, \tilde{\rho}) |_{x=0} = & - \partial_x \tTr \left( \tilde{\rho} \int_0^\infty \frac{dt}{1 + t^2}
        \frac{\tilde{\rho}_x t - 1}{\tilde{\rho}_x + t} \right) \Bigg|_{x=0} \nonumber \\
    = & \tTr \left[ \tilde{\rho} \int_0^\infty dt \frac{1}{\tilde{\rho}_0 + t} \left( \tilde{\rho}_0 - \tilde{\rho}'\right)
        \frac{1}{\tilde{\rho}_0 + t}\right] \nonumber \\
    = & 1 - \int_0^\infty dt \tTr \left( \frac{1}{\tilde{\rho}_0 + t} \tilde{\rho} \frac{1}{\tilde{\rho}_0  + t} \tilde{\rho}'
        \right). \nonumber
\end{align}
Substituting $\tilde{\rho}_0 = \sum_j \lambda_{\tau j} \frac{1}{\qd_\tau} \ket{j; \tau}_A\bra{j; \tau}
 \otimes \frac{1}{\qd_\tau} \ket{j; \tau}_B\bra{j; \tau}$ into the above equation, we have
\begin{align}
    & \frac{1}{\tilde{\rho}_0 + t} \tilde{\rho} \frac{1}{\tilde{\rho}_0  + t} \nonumber \\
    = & \sum_{j j'} \frac{\sqrt{\lambda_{\tau j} \lambda_{\tau j'}}}{(\lambda_{\tau j} \frac{1}{\qd_\tau^2} + t)
        (\lambda_{\tau j'} \frac{1}{\qd_\tau^2} + t)} \ket{j; \tau}_A \bra{j'; \tau} \nonumber \\
    & \otimes \ket{j; \tau}_B \bra{j';\tau} \otimes \ket{\tau, \tau; 1} \bra{\tau, \tau; 1}.
\end{align}
For $\lambda_{\tau j}, \lambda_{\tau j'} \in (0,1]$, we have
\begin{align}
    g(\lambda_{\tau j}, \lambda_{\tau j'}) = & \int_0^\infty dt \frac{\sqrt{\lambda_{\tau j} \lambda_{\tau j'}}}
        {(\lambda_{\tau j} \frac{1}{\qd_\tau^2} + t)(\lambda_{\tau j'} \frac{1}{\qd_\tau^2} + t)} \nonumber \\
    \le & \int_0^\infty dt \frac{\sqrt{\lambda_{\tau j} \lambda_{\tau j'}}}
        {(\sqrt{\lambda_{\tau j}\lambda_{\tau j'}} \frac{1}{\qd_\tau^2} + t)^2} \nonumber \\
    = & \qd_\tau^2.
\end{align}
Since any separable anyonic state can be decomposed by term $\frac{1}{\qd_a} \ket{\alpha; a}_A \bra{\alpha; a}
 \otimes \frac{1}{\qd_b} \ket{\beta; b}_B\bra{\beta; b}$.
We can set $\tilde{\rho}' = \frac{1}{\qd_\tau} \ket{\alpha; \tau}_A \bra{\alpha; \tau} \otimes \frac{1}{\qd_\tau}
 \ket{\beta; \tau}_B\bra{\beta; \tau}$, where $\ket{\alpha; \tau} = \sum_k a_k \ket{k; \tau}$
and $\ket{\beta; \tau} = \sum_m b_m \ket{m; \tau}$.
Then, we have
\begin{align}
    & \partial_x f(x, \tilde{\rho})|_{x=0} - 1 \nonumber \\
    = & - \int_0^\infty dt \tTr \left( \frac{1}{\tilde{\rho}_0 + t} \tilde{\rho} \frac{1}{\tilde{\rho}_0  + t}
        \tilde{\rho}' \right) \nonumber \\
    = & - \sum_{j j'} g(\lambda_{\tau j}, \lambda_{\tau j'}) \frac{1}{\qd_\tau^2} a_{j'} a^*_j b_{j'} b^*_j,
\end{align}
and
\begin{align}
    & \left| \partial_x f(x, \tilde{\rho})|_{x=0} - 1 \right| \nonumber \\
    \le & \sum_{j j'} \left| g(\lambda_{\tau j}, \lambda_{\tau j'}) \frac{1}{\qd_\tau^2} a_{j'} a^*_j b_{j'} b^*_j \right|
        \nonumber \\
    \le & \sum_{j j'} \left| a_{j'} \right| \left| a^*_j \right| \left| b_{j'} \right| \left| b^*_j \right| \nonumber \\
    = & \left( \sum_j \left| a_j \right| \left| b_j \right| \right)^2 \nonumber \\
    \le & \sum_j \left| a_j \right|^2 \sum_{j'} \left| b_{j'} \right|^2 = 1.
\end{align}
Finally, we have
\begin{equation}
    \partial_x f(x, \tilde{\rho})|_{x=0} \ge 0.
\end{equation}
We have demonstrated that $\tilde{\rho}_0$ is the separable state corresponding to the local minimum of $f(x,\tilde{\rho})$.
The proof that $\tilde{\rho}_0$ is the separable state corresponding to the global minimum value is the same
as that in Ref.~\cite{PhysRevA.57.1619},
relying on the convex nature of anyonic relative entropy~\cite{YXY}.

Next, we prove Theorem~D1.

\textbf{Proof (Theorem~D1)} --
First, let's examine the $E_{\rm ACE}$. According to Ref.~\cite{PhysRevA.108.052221} and~\cite{bonderson2017anyonic},
the $E_{\rm ACE}$ of any pure anyonic state $\tilde{\rho}$ has a general form:
\begin{equation}
    E_{\rm ACE} (\tilde{\rho}) = H\left( \left\{ p_a \right\} \right) + 2 \sum_a p_a \log_2 \qd_a, \nonumber
\end{equation}
where $H\left( \left\{ p_a \right\} \right)$ is the Shannon entropy and $p_a$ are
probabilities that satisfy $\sum_a p_a = 1$. Therefore, we have
\begin{equation}
    \lim_{n \rightarrow \infty} \frac{E_{\rm ACE}(\tilde{\rho}^n)}{n} = 0. \nonumber
\end{equation}
Furthemore, according to Eq.~(\ref{eq:Pytha}) in the main text, we have
\begin{equation}
    \lim_{n \rightarrow \infty} \frac{E_{\rm AREE} (\tilde{\rho}^n)}{n}
        = \lim_{n \rightarrow \infty} \frac{E_{\rm CE} (\tilde{\rho}^n)}{n}. \nonumber
\end{equation}
Now we only need to prove the following equality:
\begin{equation}
    \lim_{n \rightarrow \infty} \frac{E_{\rm AREE} (\tilde{\rho}^n)}{n} = E_{\rm AEE}. \nonumber
\end{equation}
For a general pure anyonic state:
\begin{equation}
    \ket{\tilde{\psi}} = \sum_{a,i} \sqrt{\lambda_{ai}} \ket{i;a}_A \ket{i;\overline{a}}_B
        \ket{a,\overline{a};1}, \nonumber
\end{equation}
the corresponding state of $n$-copy can be written as
\begin{equation}
    \ket{\tilde{\psi}}^n = \sum_{\boldsymbol{a},\boldsymbol{i}} \sqrt{\lambda_{\boldsymbol{a}\boldsymbol{i}}}
        \ket{\boldsymbol{i};\boldsymbol{a}}_A \ket{\boldsymbol{i};\overline{\boldsymbol{a}}}_B
        \ket{\boldsymbol{a},\overline{\boldsymbol{a}};1}, \nonumber
\end{equation}
where $\lambda_{\boldsymbol{a}\boldsymbol{i}} = \lambda_{a_1 i_1} \lambda_{a_2 i_2}
 \cdots \lambda_{a_n i_n}$, $\ket{\boldsymbol{i};\boldsymbol{a}}_A = \ket{i_1;a_1}_A
 \ket{i_2;a_2}_A \cdots \ket{i_n; a_n}_A$,
and $\ket{\boldsymbol{a},\overline{\boldsymbol{a}};1} = \ket{a_1,\overline{a}_1;1}
 \ket{a_2,\overline{a}_2;1} \cdots \ket{a_n,\overline{a}_n;1}$.
By using the basis transformation named the $F$-matrices
\begin{align}
     \begin{tikzpicture}[baseline, scale=0.5]
        \draw (-1,1)--(0,0) node[pos=0, above]{$a_1$}--(1,1)
          node[pos=1, above]{$\overline{a}_1$};
        \draw (-2,1)--(0,-1) node[pos=0, above]{$a_2$}--(2,1)
          node[pos=1, above]{$\overline{a}_2$};
      \end{tikzpicture}
    =  \sum_b \left[ F^{a_2 a_1 \overline{a}_1}_{a_2} \right]_{1b}
    \begin{tikzpicture}[baseline, scale=0.5]
      \draw (-1,1)--(-0.5,0.5) node[pos=0, above]{$a_1$}--(-1,0);
      \draw (-2,1)--(-1,0) node[pos=0, above]{$a_2$}--(0,-1) node[pos=0.6,above]{$b$}
        --(1,0) node[pos=0.4,above]{$\overline{b}$}--(2,1)
        node[pos=1, above]{$\overline{a}_2$};
      \draw (1,1)--(0.5,0.5) node[pos=0, above]{$\overline{a}_1$}--(1,0);
    \end{tikzpicture}, \nonumber
\end{align}
where the coefficient $\left[ F^{a_2 a_1 \overline{a}_1}_{a_2} \right]_{1b}
 = \sqrt{\qd_b/(\qd_{a_1}\qd_{a_2})}$, we can transform the
state $\ket{\boldsymbol{a},\overline{\boldsymbol{a}};1}$ into
\begin{equation}
    \ket{\boldsymbol{a},\overline{\boldsymbol{a}};1} = \sum_{\boldsymbol{b},c}
        \sqrt{\frac{\qd_c}{\qd_{\boldsymbol{a}}}}  \ket{\boldsymbol{a},\boldsymbol{b};c}
        \ket{\overline{\boldsymbol{a}},\overline{\boldsymbol{b}};\overline{c}}
        \ket{c,\overline{c}; 1}, \nonumber 
\end{equation}
where $\qd_{\boldsymbol{a}}= \qd_{a_1} \qd_{a_2} \cdots \qd_{a_n}$ and
\begin{equation}
    \ket{\boldsymbol{a},\boldsymbol{b};c} = \ket{a_1,a_2; b_1}_A 
        \cdots \ket{b_{n-2},a_n; c}. \nonumber
\end{equation}
The $n$-copy of the pure anyonic state can be rewritten as
\begin{align}
    & \ket{\tilde{\psi}}^n \nonumber \\
    = & \sum_{\boldsymbol{a}, \boldsymbol{b}, \boldsymbol{i}, c}
        \sqrt{\frac{\lambda_{\boldsymbol{a}\boldsymbol{i}} \qd_c}{\qd_{\boldsymbol{a}}}}
        \ket{\boldsymbol{i};\boldsymbol{a}}_A \ket{\boldsymbol{i};\overline{\boldsymbol{a}}}_B
        \ket{\boldsymbol{a},\boldsymbol{b};c}
        \ket{\overline{\boldsymbol{a}},\overline{\boldsymbol{b}};\overline{c}}
        \ket{c,\overline{c}; 1}. \nonumber
\end{align}
Thus, we obtain the $E_{\rm AREE}$
\begin{align}
    & E_{\rm AREE}(\ket{\tilde{\psi}^n}) \nonumber \\
    = & \sum_{\boldsymbol{a}, \boldsymbol{b}, \boldsymbol{i}, c}
        - \frac{\lambda_{\boldsymbol{a}\boldsymbol{i}} \qd_c}{\qd_{\boldsymbol{a}}} \log_2
        \frac{\lambda_{\boldsymbol{a}\boldsymbol{i}} }{\qd_{\boldsymbol{a}} \qd_c} \nonumber \\
    = & \sum_{\boldsymbol{a}, \boldsymbol{i}, c}
        - \frac{\lambda_{\boldsymbol{a}\boldsymbol{i}} \dim V_{\boldsymbol{a}}^c \qd_c}{\qd_{\boldsymbol{a}}}
        \log_2 \frac{\lambda_{\boldsymbol{a}\boldsymbol{i}} }{\qd_{\boldsymbol{a}} \qd_c} \nonumber\\
    = & \sum_{\boldsymbol{a}, \boldsymbol{i}} - \lambda_{\boldsymbol{a}\boldsymbol{i}} \log_2
        \frac{\lambda_{\boldsymbol{a}\boldsymbol{i}}}{\qd_{\boldsymbol{a}}}
        + \sum_{\boldsymbol{a}, \boldsymbol{i}, c}
        \frac{\lambda_{\boldsymbol{a}\boldsymbol{i}} \dim V_{\boldsymbol{a}}^c \qd_c}{\qd_{\boldsymbol{a}}}
        \log_2 \qd_c \nonumber \\
    = & n H\left( \left\{ \lambda_{ai} \right\} \right) + n \sum_{a,i} \lambda_{a,i} \log_2 \qd_a
        + \sum_c p_c \log_2 \qd_c, \nonumber
\end{align}
where $p_c = \sum_{\boldsymbol{a}, \boldsymbol{i}} \frac{\lambda_{\boldsymbol{a}
 \boldsymbol{i}} \dim V_{\boldsymbol{a}}^c \qd_c}{\qd_{\boldsymbol{a}}} \ge 0$.
In the second step, we have used the equality $\sum_{\boldsymbol{b}} 1
 = \dim V^c_{\boldsymbol{a}}$.
In the third step, we have used the equality $\sum_c \dim V^c_{\boldsymbol{a}}
 \qd_c = \qd_{\boldsymbol{a}}$.
Since $\sum_c p_c =1$, we obtain the asymptotic limit of the $E_{\rm AREE}$:
\begin{equation}
    \lim_{n \rightarrow \infty} \frac{E_{\rm AREE}(\ket{\tilde{\psi}^n})}{n}
        = H\left( \left\{ \lambda_{ai} \right\} \right) + \sum_{a,i} \lambda_{a,i} \log_2 \qd_a,
        \nonumber
\end{equation}
which is the same as the AEE of state $\ket{\tilde{\psi}}$.

\bibliography{FEB.bib}

\end{document}